\input harvmac.tex
\input epsf


\def\be{\begin{equation}}
\def\bea{\begin{eqnarray}}
\def\ee{\end{equation}}
\def\eea{\end{eqnarray}}
\def\d{\partial}

\def\eps{\epsilon}

\lref\FP{O.~Lunin and S.~D.~Mathur,
``Metric of the multiply wound rotating string,''
Nucl.\ Phys.\ B {\bf 610}, 49 (2001)
[arXiv:hep-th/0105136].}

\lref\supertubes{D.~Mateos and P.~K.~Townsend,
``Supertubes,''
Phys.\ Rev.\ Lett.\  {\bf 87}, 011602 (2001)
[arXiv:hep-th/0103030]\semi
R.~Emparan, D.~Mateos and
P.~K.~Townsend, ``Supergravity supertubes,'' JHEP {\bf 0107}
(2001) 011 [arXiv:hep-th/0106012]\semi
D.~Mateos, S.~Ng and
P.~K.~Townsend, ``Tachyons, supertubes and brane/anti-brane
systems,'' JHEP {\bf 0203}, 016 (2002) [arXiv:hep-th/0112054]\semi
D.~Mateos, S.~Ng and P.~K.~Townsend,
``Supercurves,'' Phys.\ Lett.\ B {\bf 538}, 366 (2002)
[arXiv:hep-th/0204062].}

\lref\chonam{
J.~H.~Cho and S.~k.~Nam,
``AdS(3) black hole entropy and the spectral flow on the horizon,''
arXiv:hep-th/9903058.
}

\lref\MCP{C.~G.~Callan, J.~M.~Maldacena and A.~W.~Peet,
``Extremal Black Holes As Fundamental Strings,'' Nucl.\ Phys.\ B
{\bf 475}, 645 (1996) [hep-th/9510134].}

\lref\Harvey{A.~Dabholkar, J.~P.~Gauntlett, J.~A.~Harvey and
D.~Waldram, ``Strings as Solitons and Black Holes as Strings,''
Nucl.\ Phys.\ B {\bf 474}, 85 (1996) [hep-th/9511053].}

\lref\MalRusso{J.~M.~Maldacena and J.~G.~Russo, ``Large N limit
of non-commutative gauge theories,'' JHEP {\bf 9909}, 025 (1999)
[arXiv:hep-th/9908134].}

\lref\DDbar{ D.~s.~Bak and A.~Karch, ``Supersymmetric
brane-antibrane configurations,'' Nucl.\ Phys.\ B {\bf 626}, 165
(2002) [arXiv:hep-th/0110039]\semi
A.~R.~Lugo, ``On supersymmetric Dp anti-Dp brane
solutions,'' Phys.\ Lett.\ B {\bf 539}, 143 (2002)
[arXiv:hep-th/0206041].}

\lref\myerslight{
R.~C.~Myers and D.~J.~Winters,
``From D - anti-D pairs to branes in motion,''
arXiv:hep-th/0211042.
}

\lref\aich{
P.~C.~Aichelburg and R.~U.~Sexl,
``On The Gravitational Field Of A Massless Particle,''
Gen.\ Rel.\ Grav.\  {\bf 2}, 303 (1971).
}

\lref\bachas{
C.~Bachas and C.~Hull,
``Null brane intersections,''
arXiv:hep-th/0210269.
}
\lref\vafagas{
C.~Vafa,
``Gas of D-Branes and Hagedorn Density of BPS States,''
Nucl.\ Phys.\ B {\bf 463}, 415 (1996)
[arXiv:hep-th/9511088].
}

\lref\CNM{ G.~T.~Horowitz and A.~A.~Tseytlin, ``A New class of
exact solutions in string theory,'' Phys.\ Rev.\ D {\bf 51}, 2896
(1995) [arXiv:hep-th/9409021] \semi
A.~A.~Tseytlin, ``Generalised
chiral null models and rotating string backgrounds,'' Phys.\
Lett.\ B {\bf 381}, 73 (1996) [arXiv:hep-th/9603099].
}

\lref\Gradshteyn{ I.~S.~Gradshteyn and I.~M.~Ryzhik, ``Table of
Integrals, Series and Products,'' ,San Diego : Academic Press,
2000 , Chapter 8.}

\lref\desing{J.~M.~Maldacena and L.~Maoz, ``De-singularization by
rotation,'' arXiv: hep-th/0012025.}

\lref\IWP{Z.~Perjes, ``Solutions Of The Coupled Einstein Maxwell
Equations Representing The Fields Of Spinning Sources,'' Phys.\
Rev.\ Lett.\  {\bf 27}, 1668 (1971) \semi W.~Israel and G.~Wilson,
J.\ Math.\ Phys.\ {\bf 13} (1972) 865.}

\lref\lmRot{
O.~Lunin and S.~D.~Mathur,
``Rotating deformations of AdS(3) x S**3, the orbifold CFT
and strings in  the pp-wave limit,''
Nucl.\ Phys.\ B {\bf 642}, 91 (2002)
[arXiv:hep-th/0206107].
}

\lref\lmBH{
O.~Lunin and S.~D.~Mathur,
``AdS/CFT duality and the black hole information paradox,''
Nucl.\ Phys.\ B {\bf 623}, 342 (2002)
[arXiv:hep-th/0109154].
}

\lref\lms{
O.~Lunin, S.~D.~Mathur and A.~Saxena,
``What is the gravity dual of a chiral primary?,''
arXiv:hep-th/0211292.
}

\lref\deboeretal{
V.~Balasubramanian, J.~de Boer, E.~Keski-Vakkuri and S.~F.~Ross,
``Supersymmetric conical defects: Towards a string theoretic description  of black hole formation,''
Phys.\ Rev.\ D {\bf 64}, 064011 (2001)
[arXiv:hep-th/0011217].
}

\lref\wadia{
J.~R.~David, G.~Mandal, S.~Vaidya and S.~R.~Wadia,
``Point mass geometries, spectral flow and AdS(3)-CFT(2) correspondence,''
Nucl.\ Phys.\ B {\bf 564}, 128 (2000)
[arXiv:hep-th/9906112].
}

\lref\cvetic{ M.~Cvetic and D.~Youm, ``General Rotating Five
Dimensional Black Holes of Toroidally Compactified Heterotic
String,'' Nucl.\ Phys.\ B {\bf 476}, 118 (1996)
[arXiv:hep-th/9603100] \semi
M.~Cvetic and F.~Larsen,
``Near horizon geometry of rotating black holes in five dimensions,''
Nucl.\ Phys.\ B {\bf 531}, 239 (1998)
[arXiv:hep-th/9805097].
}

\lref\jmas{
J.~M.~Maldacena and A.~Strominger,
``AdS(3) black holes and a stringy exclusion principle,''
JHEP {\bf 9812}, 005 (1998)
[arXiv:hep-th/9804085].
}
\lref\deboercp{
J.~de Boer,
``Six-dimensional supergravity on S**3 x AdS(3) and 2d conformal field  theory,''
Nucl.\ Phys.\ B {\bf 548}, 139 (1999)
[arXiv:hep-th/9806104].
}

\lref\sezgincp{
S.~Deger, A.~Kaya, E.~Sezgin and P.~Sundell,
``Spectrum of D = 6, N = 4b supergravity on AdS(3) x S(3),''
Nucl.\ Phys.\ B {\bf 536}, 110 (1998)
[arXiv:hep-th/9804166].
}

\lref\seibergwitten{
N.~Seiberg and E.~Witten,
``The D1/D5 system and singular CFT,''
JHEP {\bf 9904}, 017 (1999)
[arXiv:hep-th/9903224].
}

\lref\review{
O.~Aharony, S.~S.~Gubser, J.~M.~Maldacena, H.~Ooguri and Y.~Oz,
``Large N field theories, string theory and gravity,''
Phys.\ Rept.\  {\bf 323}, 183 (2000)
[arXiv:hep-th/9905111].
}

\lref\mcelgintwo{
E.~J.~Martinec and W.~McElgin,
``Exciting AdS orbifolds,''
JHEP {\bf 0210}, 050 (2002)
[arXiv:hep-th/0206175].
}

\lref\mcelginone{
E.~J.~Martinec and W.~McElgin,
``String theory on AdS orbifolds,''
JHEP {\bf 0204}, 029 (2002)
[arXiv:hep-th/0106171].
}

\lref\antibranes{
D.~s.~Bak and A.~Karch,
``Supersymmetric brane-antibrane configurations,''
Nucl.\ Phys.\ B {\bf 626}, 165 (2002)
[arXiv:hep-th/0110039].
}
\lref\citePP{
J.~Kowalski-Glikman,
``Vacuum States In Supersymmetric Kaluza-Klein Theory,''
Phys.\ Lett.\ B {\bf 134}, 194 (1984)\semi
J.~Figueroa-O'Farrill and G.~Papadopoulos,
``Homogeneous fluxes, branes and a maximally supersymmetric
solution of M-theory,''
JHEP {\bf 0108}, 036 (2001)
[arXiv:hep-th/0105308]\semi
M.~Blau, J.~Figueroa-O'Farrill, C.~Hull and G.~Papadopoulos,
``Penrose limits and maximal supersymmetry,''
Class.\ Quant.\ Grav.\  {\bf 19}, L87 (2002)
[arXiv:hep-th/0201081]\semi
D.~Berenstein, J.~M.~Maldacena and H.~Nastase,
JHEP {\bf 0204}, 013 (2002)
[arXiv:hep-th/0202021].
}

\lref\Lugo{N.~E.~Grandi and A.~R.~Lugo, ``Supertubes and special
holonomy,'' arXiv:hep-th/0212159.
}


{ \Title{\vbox{\baselineskip12pt \hbox{hep-th/0212210} }} {\vbox{
{\centerline { Gravity solutions for the D1-D5 system  }}
{\centerline{  with angular momentum}}
}}}
\bigskip
\centerline{ Oleg Lunin$^{1}$, Juan Maldacena$^{1}$ and Liat
Maoz$^{2,3}$ }
\bigskip
\centerline{$^1$ Institute for Advanced Study, Princeton, NJ 08540,USA}
\bigskip
\centerline{$^2$ Jefferson Physical Laboratory, Cambridge, MA 02138,
USA}
\bigskip
\centerline{$^3$ Jadwin Hall, Princeton, NJ 08544,USA}

\vskip .3in

We construct a large family of
 supergravity solutions that describe BPS excitations on
$AdS_3 \times S^3$ with angular momentum on $S^3$. These solutions
take into account the full backreaction on the metric.
 We find that as we increase the energy of the excitation,
the energy gap to the next non-BPS excitation decreases.
These solutions can be viewed as Kaluza-Klein monopole
``supertubes'' which
are completely non-singular geometries.
We also make some remarks on supertubes in general.


\newsec{Introduction}

Perhaps one of the most distinctive aspects of gravity is that
time slows down near heavy objects due to gravitational redshift.

We now have many cases where we have dual descriptions of gravitational theories in terms of ordinary quantum
field theories via the AdS/CFT correspondence.  It is interesting then to find situations where this effect is
under some degree of control, so that we can understand it from the field theory point of view. Black holes are
extreme examples where this redshift factor goes to zero. In this paper we consider configurations where this
redshift factor is important but does not go to zero.

We focus on $AdS_3 \times S^3$ compactifications, and we consider states with angular momentum on $S^3$ that are
BPS. These states are also called ``chiral primary'' states. When these states carry large amounts of angular
momentum, their back-reaction on the metric cannot be ignored. In this paper we construct exact gravity solutions
which take this backreaction into account. We indeed find that there is an important redshift effect that implies,
among other things, that the energy gap to the next non-BPS excitation decreases as we increase the angular
momentum. This gap goes to zero for certain states that are on the verge of forming black holes.

These solutions can be found by noticing that the D1/D5 system with angular momentum blows up into a Kaluza-Klein
monopole supertube, U-dual to the one described in \supertubes .  Since the Kaluza-Klein monopole is non-singular,
these geometries are non-singular. The configuration with maximal angular momentum, which corresponds to a
supertube with circular shape, has a near horizon geometry equal to $AdS_3 \times S^3$ in global coordinates.
Supertubes with non-circular shapes correspond to chiral primary excitations on the $AdS_3 \times S^3$ vacuum.

The solutions are also interesting since they provide non-singular gravity solutions for configurations that are
1/4 BPS in toroidally compactified string theory. Different gravity solutions are related to different microscopic
states.

Previous work on the subject focused on gravity solutions with conical singularities. We show that these conical
singularities are not a good description of the long distance properties of generic chiral primaries, i.e. the
non-singular solutions are different, even at long distances. Some very special chiral primaries can give conical
metrics with opening angles of the form $2\pi/N$. Conical metrics with non-integer angles are not a good
approximation to any of the non-singular metrics. Singular geometries more closely related to chiral primaries can
be found in  \lms . We will  show that our solutions look like the solutions in \lms\ at long distances.

In this paper we also analyze some aspects of the geometry of
supertubes in other dimensions and in various limits.

This paper is organized as follows. In section two we describe the construction of the gravity solutions. In
section three we discuss the relation of these gravity solutions to the problem of chiral primaries in $AdS_3
\times S^3$. In section four we describe general non-singular solutions with plane wave asymptotic boundary
conditions, which can be thought of as arising from particles propagating on plane wave backgrounds. In section
five we discuss some aspects of the gravitational geometry  of supertubes in various dimensions. This section is a
bit disconnected from the previous part of the paper.

\newsec{ Gravity solutions for the D1-D5 system with angular momentum}

In this section we consider ten dimensional supergravity compactified on $S^1 \times T^4$,
 and consider a
system of $Q_1$ D1 branes wrapped on $S^1$ and $Q_5$ D5 branes wrapped on all the compact directions\foot{In
appendix B we explain how to obtain the solutions for the $K3$ case. }. We are interested in constructing
solutions which carry angular momentum and are 1/4 BPS. In other words, they are as BPS as the D1 and D5 branes
with no angular momentum. Since there are four non-compact transverse directions, the angular momenta live in
$SO(4) \sim SU(2)_L \times SU(2)_R$. The angular momentum is bounded by $J_L , J_R \leq  k \equiv Q_1 Q_5 $
\vafagas .
 For large values of $Q$ the
angular momentum can be macroscopic and can have an important effect on the geometry of the configuration. This
was initially explored in \desing \deboeretal\ who found that the geometry with maximal angular momentum was
non-singular. In the meantime, studies of other 1/4 BPS configurations with angular momentum  have given rather
interesting results. The best known example is the so called ``supertube'' which is a configuration carrying D0
and fundamental string charges with angular momentum, which is described in terms of a tubular D2 brane with
electric and magnetic fields on its worldvolume \supertubes .
 The configuration with maximal angular momentum consists of a
 tubular D2 brane with a radius square proportional to the product of the two charges. The configuration
does not carry any net D2 brane charge. Tubes with arbitrary
cross sections are also possible, but they carry less angular
momentum \supertubes .

The D0-F1 system is U-dual to D1-D5.
Under this
U-duality the above D2 brane goes over to a Kaluza Klein monopole
which is wrapped on $T^4$ and a circle in the four non-compact
dimensions. The special circle of the KK monopole is the $S^1$ common
to the D1 and D5 branes. The  gravity solution for
a circular KK monopole was found in \refs{\desing,\deboeretal} (based
on the general solutions in \cvetic ) though it was
not given this description (which is not that obvious by just looking
at the metric). This solution is non-singular
because the KK monopole has a non-singular
geometry.

Now we construct similar solutions with arbitrary shapes which
are also non-singular. The technique we use to find the solution
is based on the observation that this system is U-dual to
fundamental strings with momentum along the string. Microscopic
configurations of the system are given by strings carrying
traveling waves along them, in other words, strings with only
left (or only right) moving excitations. For this case there are
gravity solutions that closely correspond  to given microscopic
states \refs{\MCP,\Harvey} Namely these solutions describe an
oscillating string with an arbitrary profile ${\bf F}(v)$, where
$v$ is a lightcone coordinate along the string.  By a chain of
dualities these can be mapped to the D1-D5 system
 so that we find the solution \lmBH~ (see appendix B)
\eqn\DoDfChiral{\eqalign{ ds^2&=f_1^{-1/2} f_5^{-1/2} \left[-(dt-A_idx^i)^2+(dy+B_idx^i)^2\right] + f_1^{1/2}
f_5^{1/2} d{\bf x}\cdot d{\bf x} \cr &+ f_1^{1/2} f_5^{-1/2} d{\bf z}\cdot
d{\bf z}\cr e^{2\Phi} &= f_1
f_5^{-1}, \cr \qquad C^{(2)}_{ti}= & { B_i\over f_1} ,\qquad C^{(2)}_{ty}=f_1^{-1} - 1 , \qquad \cr
 C^{(2)}_{iy} = &
-{A_i\over f_1},\qquad C^{(2)}_{ij}={\cal C}_{ij}+ f_1^{-1}
(A_iB_j-A_jB_i) }} The functions $f_{1,5}$ and $A_i$ appearing in
this solution are related to the profile ${\bf F}(v)$
\eqn\HarmFunc{ f_5=1+{Q_5\over L}\int_0^L{dv\over|{\bf x}-{{\bf
F}}|^2},\quad f_1=1+ { Q_5\over L}\int_0^L{|\dot F|^2dv\over
|{\bf x}-{{\bf F}}|^2},\quad A_i=-{Q_5\over L}\int_0^L{{\dot
F}_i dv\over|{\bf x}-{{\bf F}}|^2} }
 and the forms $B_i$ and ${\cal C}_{ij}$ are defined
by the duality relations
 \eqn\DualFields{ d{\cal C}=-*_4 df_5,\qquad
dB=- *_4dA. } where the $*_4$ is defined in the four non-compact spatial dimensions. The one brane
charge is given by \eqn\onebr{ Q_1 = Q_5 \langle |\dot F|^2 \rangle = Q_5 { 1 \over L} \int_0^L{|\dot F|^2 dv} }
The length $L$ that appears in these formulas is
\eqn\LDoDf{ L ={ 2\pi n_5 \over R} = 2\pi n_5 R'
   } where $n_5$ is the number of fivebranes and $R$ is the radius of the $y$ circle, while $R'$
is the radius in the original fundamental string description\foot{ For simplicity we have set $g = \alpha' = V_4 =
1$ in the above formulas. In that case $Q_1 = n_1$ and $Q_5 = n_5$, otherwise $Q_i$ have dimensions of length
square and denote the contribution of the onebranes and fivebranes to the gravitational radius of the
configuration, while $n_{1,5}$ are integers. }. $n_5$ is the original number of strings which becomes the number
of fivebranes. We see that we are taking a configuration where the string is multiply wound. This will be
important for later considerations. Configurations where the string consists of independent pieces can be obtained
by adding the corresponding contributions in the harmonic functions \HarmFunc . The solutions are parameterized by
the profile ${\bf F}(v)$ which describes a trajectory in the four non-compact dimensions. Note that the final
solution \DoDfChiral\ is time independent. We will see that the $v$ dependence of $F$ translates into a dependence
of the solution on the non-compact dimensions. The angular momentum of the solution \DoDfChiral\ is given by
\eqn\angmom{ J_{ij}={Q_5 R \over L} \int_0^L  (F_i \dot F_j - F_j \dot F_i) dv } It can be checked that the
angular momentum is always smaller than $n_1 n_5$. We will later concentrate on the two $U(1)$ components $J_\phi
= J_{12}$, $J_\psi = J_{34}$ and define $2 J_{L,R} = J_\phi \pm J_\psi$.

Note that all these solutions correspond to different ground
states of the D1/D5 system. This system has a large degeneracy, of
order $e^{ 2 \pi \sqrt{ 2 Q_1Q_5} }$.\foot{More precisely, these
solutions are particular combinations of states of the theory.
Classically there is an infinite number of solutions since they
are parameterized by continuous parameters. In the quantum theory
we should quantize the moduli space of solutions and that will
give us a finite number. This quantization is expected to give us
the same as quantizing the left movers on a string, though we did
not verify it explicitly.}.

\subsec{ An argument showing the solution is non-singular }

Looking at the metric \DoDfChiral, one might think that it is singular if ${\bf x}={\bf F}(v_0)$ for some value of
$v_0$, since the harmonic functions \HarmFunc\ diverge there. However, it was shown in \refs{\desing,\deboeretal}
that  the maximally rotating solution is non-singular. The maximally rotating solution corresponds to a
circular profile
 \eqn\eqa{
F_1 + i F_2 = a e^{ i \omega v}
\qquad F_3=F_4=0 , ~~~~~{\rm with }~~~ \omega = { 2 \pi \over L} =
{R \over n_5} }
 {}From the expression for the charges \onebr\ we get that the radius
is
\eqn\MaxA{
 a={\sqrt{Q_1Q_5}\over R}
}
On the other hand if we have a circular profile with a frequency
$\omega' = n \omega$ (and $a' = a/n$), we would
get a geometry which has a conical singularity of opening angle
$2\pi/n$.

Let us now look at the geometry corresponding to a more general
profile $F(v)$. We will analyze the metric near the potential
singularity ${\bf x}={\bf F}(v_0)$ and show that for a generic
profile ${\bf F}(v)$ the solution is completely regular. By
generic we mean a profile satisfying two conditions:

(i) the profile does not have self--intersections (if $v_1\ne
v_2$, then ${\bf F}(v_1)\ne {\bf F}(v_2)$);

(ii) the derivative ${\dot{\bf F}}(v)$ never vanishes.

Looking at the vicinity of the singularity for such profile, we
find
 \eqn\eqd{\eqalign{ f_5 &\approx {Q_5\over L}\int_{-L/2}^{L/2}{dv\over
|{\bf x}-{\bf F}|^2}\approx {Q_5\over L}\int_{-L/2}^{L/2}{dv\over x_\perp^2+({\dot F})^2 v^2}={Q_5\over
L}{\pi\over |\dot F|x_\perp}, \cr f_1 &\approx  {Q_5\over L}{\pi|\dot F| \over x_\perp},\qquad A_i\approx
-{Q_5\over L}{\pi{\dot F}_i \over |\dot F|x_\perp} }} We have  split the coordinates of the transverse space
around the point ${\bf F}(v_0) $ into a longitudinal piece, $x_l$,
 along $\dot F(v_0)$ and
a transverse piece $x_\perp$.

The asymptotics \eqd\  can be used to show that there are no
singularities in the longitudinal piece of the metric
\eqn\LongMetr{ ds_l^2\equiv |\dot F| \left[
f_5  dx_l^2- f_1^{-1} |A_i|^2
dx_l^2 \right ], }
 but they are not good enough for finding the finite
contribution to $ds_l$. We will refer to the appendix F of \lmBH\
where more careful analysis was done, and give the result
\eqn\eqe{ ds_l^2=|\dot F| Cdx_l^2 } where $C$ is a positive numerical
coefficient whose value depends on {\it global} properties of the
profile
 \eqn\eqf{ C(v_0)= {1\over |{\dot{\bf
F}}(v_0)|^2}\left\{ {Q_5\over L}\int_0^{L} {dv({\dot{\bf
F}}(v)-{\dot {\bf F}}(v_0))^2\over ({{\bf F}}(v)-{ {\bf
F}}(v_0))^2} + (1+|{\dot{\bf{F}}}(v_0)|^2)\right\} }
Let us now analyze the metric in the space transverse to the
singularity
 \eqn\eqg{ ds_\perp^2\equiv |\dot F| \left[
f_5(dx_\perp^2+x_\perp^2d\Omega^2_2)+ f_1^{-1} (B_idx^i)^2
\right]}
In order to compute the leading order terms in the metric it is
important to compute $B_i$ which is dual to $A_i$. We only need to
compute this to leading order in $x_\perp$ so that we find
\eqn\eqj{
 B_\psi \sim -(\cos\theta-1)
{\pi Q_5\over L} } where the metric in the flat transverse space
is
parameterized as
\eqn\eqi{
ds_0^2=dx_l^2+dx_\perp^2+x_\perp^2(d\theta^2+\sin^2\theta
d\psi^2) } Note that the range of $\theta$ is $0\le \theta<\pi$.
%
Then the transverse metric \eqg\ becomes
\eqn\eqk{
ds_\perp^2= {4Q_5 \pi \over L}
\left[(d\sqrt{x_\perp})^2+x_\perp \left\{\left({d\theta\over
2}\right)^2+\sin^2{\theta\over 2}d\psi^2\right\}\right]
} Let us
now look at the complete metric
 \eqn\eql{
ds^2=|\dot F| Cdx_l^2+ds^2_\perp+{Lx_\perp\over \pi Q_5}
\left\{dy^2+2B_idx^idy-dt^2+2A_idx^idt\right\}, } Near the
singularity we get
\eqn\RegMtro{\eqalign{
ds^2={4Q_5\pi \over
L} &\{(d\sqrt{x_\perp})^2+\cr &+x_\perp
\left[\left({d\theta\over 2}\right)^2+\sin^2{\theta\over
2}\left(d\psi+ {dy\over R} \right)^2 +
\cos^2{\theta\over 2}{dy^2\over R^2 } \right]\}\cr &+|\dot F|
C(dx_l-{1\over
C|\dot F|}dt)^2-{1\over C|\dot F|}dt^2
}}
where we used \LDoDf .
 Let us introduce new
coordinates
\eqn\GeneralChi{ \chi= {y \over R} ,\quad \quad
{\tilde\psi}=\psi+\chi,\quad \quad {\tilde\theta}={\theta\over 2},\quad
\quad \rho=\sqrt{x_\perp}
}
We see  that $\chi={y\over R}$ has periodicity $2\pi$, and
thus the change of coordinates from $\psi$ to ${\tilde\psi}$ is
well defined.  We also note that $\tilde\theta$ has a range $0\le
{\tilde\theta}<\pi/2$ and the metric \RegMtro\ becomes:
\eqn\RegMtrt{\eqalign{
ds^2&={4Q_5\pi \over L}\left\{d\rho^2+\rho^2
\left[d{\tilde\theta}^2+\sin^2{\tilde\theta}d{\tilde\psi}^2
+\cos^2{\tilde\theta}d\chi^2\right]\right\}\cr &+
|\dot F|C(dx_l-{1\over
C|\dot F|}dt)^2-{1\over C|\dot F|}dt^2
}}
 The first line in the
above expression gives the metric of a flat four dimensional space
from which we conclude that
the geometry is regular near the string profile. Note also that
the Killing vector $\partial_t $  becomes light like at
$\rho = 0$.

To summarize, what is happening is that the circle $y$ is shrinking to zero size as we approach the string source
but the presence of the field $B_i$ implies that it is non-trivially fibered over the $S^2$ that is transverse to
the line described by ${\bf F}$ in $R^4$ and we see that the $y$ circle
combines with the two sphere and $x_\perp$
to give a non-singular space, precisely as it happens for the Kaluza Klein monopole.
Note also that
a  non-trivial condition on the harmonic functions characterizing the
solution arises  from demanding that the
$B_i$ field leads to a well defined fibration of the $y$ circle on the
four dimensional space away from the sources.
This quantization condition on $B_i$ translates into
a quantization condition for the field $A_i$, which is obeyed with
a unit coefficient if we take and $A_i$ field as in \eqd\ with
$2 \pi Q_5 /L = R_y$.

In fact we can view these metrics as ``supertubes'' analogous to the ones described in \supertubes\ where the
D1-D5 system blows up to a KK monopole that wraps the four directions of $T^4$ and a curve of shape given by the
profile ${\bf F}(v)$ in the non-compact directions. The circle of the KK
monopole is the $y$ circle where the
initial D1 and D5 are wrapped. The final geometry is non-singular if the system blows up into a single KK
monopole. This is ensured if the
 function ${\bf F}$ does not self intersect.

One would like to know what the topology of these solutions is. We can ignore the $T^4$ for this question. Before
we put the D1-D5 system, the topology of the six dimensional space is $R\times R^4 \times S^1$. The topology of a
fixed radius surface far away is $ R \times S^3 \times S^1$. When we go in the radial direction the $S^3 $ is
filled in so that we have $R^4$. Let us start with the non-singular maximally rotating  circular solution. The
topology of a fixed radius surface far away is the same as above but when we go in we fill in the $S^1$ so that
the final topology is $R\times R^2 \times S^3 $. This is shown explicitly in appendix A .
 The topology for geometries that can be obtained as
continuous deformations of the circle is still going to be the same. The actual metric and geometry of the
solution will of course depend on many parameters. For example in the regime that $ a^2 \gg  Q_1 , Q_5$
 we find that the gravity
solution has the shape of  a ring whose gravitational thickness
(of order $\sqrt{Q_1}$, or $\sqrt{Q_5}$)
is much smaller than its radius.

An important lesson is that these configurations with
D-brane charge can change the topology of the spacetime where they
live. This situation is common in many examples of AdS/CFT.
It is an example of a so called ``geometric transition''.

\subsec{Geometries with $A_N$ singularities}

There are some singular geometries that are believed to be allowed
in string theory. A particular example arises when we
multiply each non-constant piece in the harmonic functions \HarmFunc\
by $N$. In that case we get a $Z_N$ singularity on the ring.
In other words, instead of a single KK monopole we have $N$ coincident
ones giving rise to an $A_{N-1}$
 singularity.
The resulting geometry is a $Z_N$ quotient
of the geometries we discussed in the previous subsection. It is easy to see that we can find non-singular
deformations of these geometries by separating
the $N$ copies of the harmonic functions in the transverse
directions. When this separation is small we get that the $A_{N-1}$ singularity becomes locally a smooth ALE
space. This situation was explored in
 detail in \mcelgintwo\ and we refer the reader to it for the details.
Notice that if $n_1$ and $n_5$ are coprime and we are at a generic point in moduli space \seibergwitten , then it
is not possible to deform the $A_{N-1}$ singularity by displacing the ``center of mass'' of the rings in the
non-compact directions, but it is possible to deform it by combining the whole system into a single string. When
we start with $N$ coincident rings as in \eqa\ we get a geometry that is a $Z_N$ quotient of $AdS_3 \times S^3$.
This has fixed points along an equator of $S^3$ and the origin of $AdS_3$. The metric can be written as the
conical metric in Appendix C, with a coefficient $\gamma^{-1} = N$. Conical metrics with arbitrary non-integer
$\gamma^{-1}$ have been considered in the literature. These metrics are suspicious since they would correspond to
KK monopoles which do not obey the proper quantization condition. Indeed, if one looks at those metrics one finds
that there are extra conical singularities as compared to the $A_N$ case. These are easy to understand in the
space parameterized by the flat coordinates $x^i$. In this case we have ``Dirac-strings'' coming out of the KK
monopole extended along the disk in the 12 plane with boundary on the ring. If the quantization condition is not
obeyed, then the metric will be singular on this disk. Since we found a large family of non-singular metrics one
might wonder if one could take a smooth metric which approximates these conical spaces with $\gamma^{-1}$ not
integer arbitrarily well. We argue in appendix C that this is not possible. In conclusion, conical metrics with
$\gamma^{-1}$ not integer are not a good approximation to the real solutions.

\newsec{Geometries corresponding to chiral primaries.}

We first need to take the decoupling limit of the solutions that
we considered above \DoDfChiral .
 This amounts to dropping the ones in the
harmonic functions $f_1$ and $f_5$ in \HarmFunc .
We can then see that the asymptotic geometry for large $|x|$ is that
of $AdS_3\times S^3$. If we take the standard periodic conditions
on the spinors along
 $S^1$ that preserve supersymmetry in the asymptotically flat context then
we see that we are in the Ramond sector of the theory.
Different solutions correspond to different Ramond vacua. These
vacua can have  various values of angular momenta ranging over
 $ - k/2 \leq J_{L,R} \leq k/2 $.
The solution with a circular profile in the $12$ plane
 corresponds to the Ramond
vacuum with maximal angular momentum $J_L = J_R = k/2$.
Under spectral flow this state goes over to the NS vacuum.
Spectral flow in the CFT is an operation that maps states in the
R sector to states in the NS sector. It is a rather trivial operation
involving only the overall U(1) R-charge of the states so we do
not expect a significant change in the properties of the state
when we perform it. In fact  spectral flow amounts to
a simple coordinate redefinition
\eqn\coordred{
 \tilde \phi = \phi - {t\over R}
~~~~~\tilde \psi = \psi + {y\over R} } with these new variables the time independent configurations that we had in
the Ramond sector can become time dependent if they depended on $\phi$. For example, the $\phi$ independent ring
solution \eqa\ becomes the time independent $AdS_3 \times S^3 $ vacuum. Ramond sector solutions where the ring is
deformed to other shapes, like an ellipse, for example,  become time dependent when they are viewed as NS sector
solutions. This is related to  the fact that chiral primaries carry non-zero energy in the NS sector. Under
spectral flow all RR vacua correspond to chiral primary states with \eqn\nscharg{\eqalign{ J^{NS}_{L,R} =&
J_{L,R}^R - { k \over 2} \cr L_0^{NS} =& |J^{NS}_R| ~,~~~~~~~~~~ \bar L^{NS}_0 = |J^{NS}_L| ~.}} The physical
properties of the solution in the interior of the space do not change when we do spectral flow, since it is just a
redefinition of what we mean by energy and angular momentum\foot{ Note, in particular, that spectral flow acting
on the NS vacuum {\it does not} produce conical singularities as was asserted in \chonam \wadia\ . A more detailed
discussion about the action of spectral flow can be found in \deboeretal \desing .}. Note, in particular, that the
statement that the $M=0$ BTZ black hole is the Ramond ground state is  imprecise. There are many Ramond
ground states and they look quite different in the supergravity description depending on their angular momenta.

As discussed in \jmas \deboercp \sezgincp\
chiral primary states close to the
NS ground state can be obtained by adding perturbative gravity
modes on the NS ground state. They are particular gravity modes
that are BPS.
First we restrict to deformations of the ring into a more general shape
in the 12 plane. There are two classes of deformations we can
consider. One is a change in the shape of the ring and the
other is small changes in the velocity with which we go around the
ring. These two correspond to two towers of chiral primary states.

The supergravity chiral primary states are given as follows
\jmas \deboercp \sezgincp .
It is convenient to separate the three form field strengths in six
dimensions
into self dual and anti-self dual parts. The background fields
in $AdS_3 \times S^3$ are self dual. The chiral primary fields correspond to fluctuations
in the anti-self dual part of the three form field strengths on the
$S^3$ (which also mix with fluctuations of scalar fields).
These fields produce gravity modes with $(J_L, J_R) = 1/2, 1, 3/2,
\cdots$. There is also one special tower of supergravity fields
which starts at $(J_L, J_R) =  1, 3/2 , \cdots $. These come from
certain fluctuations in the metric of the three sphere.

The two classes of deformations of the ring that we discussed
above correspond to two of these towers of chiral primaries.
More precisely, changes in the velocity correspond to the tower
associated to the anti-self dual component of the field strength
whose self dual component
 is turned on in the background. The changes in shape
correspond to the tower associated with deformations of the sphere. This can be seen by noticing that the lowest
angular momentum deformation of the shape we can do is to deform the circle to an ellipse. This has angular
momentum $J_\phi =2$ which corresponds to $J_L = J_R =1$. On the other hand we can change the velocity by $ v \to
v + \epsilon \cos v$ which will introduce a mode with angular momentum $J_\phi = 1$ which corresponds to $(J_L,
J_R) = (1/2, 1/2)$. In summary, different Fourier modes in the expansion of $F_{1,2}$ around the circular profile
are in direct correspondence with chiral primary gravity modes with different values of angular momentum on $S^3$
(i.e. different values of $J_\phi$).

There is another chiral primary tower  with $(J_L, J_R) =
(m+1,m), ~~m=0,1/2, \dots$ and a similar one with $J_L
\leftrightarrow J_R$. These corresponds to oscillations of the
ring into the 34 plane.

Finally we should consider many other chiral primaries that come
from the  anti-self dual components of other field strengths.
These are easily described in the $T^4$ theory as oscillations in
the internal $T^4$ directions of the initial fundamental string
which we used to construct the solutions. We give the general
solutions for those in appendix B. One nice aspect of those
solutions is that we can easily find some solutions for which the
metric is $\phi$ independent. For example, choosing a simple
profile where the string is also oscillating with frequency
$m\omega$ in the internal torus, we obtain in the
near-horizon-limit the six dimensional metric (see appendix B)
\eqn\internal{\eqalign{ ds^2\over \sqrt{Q_1 Q_5}&=(r^2+\beta
\cos^2\theta) {1 \over\sqrt{\alpha}}\left[-(dt-{\beta \sin^2\theta
d\phi\over r^2+\beta \cos^2\theta})^2+ (d\chi+{\beta \cos^2\theta
d\psi\over r^2+\beta \cos^2\theta})^2\right]\cr
&+{\sqrt{\alpha}dr^2\over r^2+\beta }+\sqrt{\alpha}d\theta^2+
{\sqrt{\alpha}\over r^2+\beta \cos^2\theta}(r^2\cos^2\theta
d\psi^2+ (r^2+\beta)\sin^2\theta d\phi^2) }} where the function
$\alpha$ is given by \eqn\ExTorNH{ \alpha =  1- (1 - \beta)
 \left({\beta \sin^2\theta\over {r^2+\beta}}\right)^{m}
} and $\beta$ and $m$ are two parameters characterizing the
solution. $m$ is  the angular momentum of the single particle
chiral primary we are exciting with $(J_L,J_R) = ({ m\over 2},{
m\over 2})$. In other words we are considering a coherent state
associated to this single particle chiral primary. The parameter
$ 0 \leq \beta  \leq 1$ measures the total angular momentum of
the solution which is \eqn\nsangmom{ J_L^{NS}=J_R^{NS}  =
{n_1n_5\over 2}(1- \beta ) } So that for $\beta =1$ we get global
$AdS_3 \times S^3$ and for $\beta=0$ we get the singular geometry
corresponding to the
 $M=0$ BTZ black hole
which
from the NS point of view could be described as the extremal limit
of a black hole that is rotating in the internal $S^3$.
For $\beta >0$ the geometry is non-singular.
We can ask if the metric \internal\ goes over to the metric of
a conical defect. Since the angular momentum is given by \nsangmom\
we would expect that the opening angle of the corresponding
conical defect should be $2 \pi \beta$, or $\gamma = \beta$ in the
notation of appendix C.
 These conical metrics were considered in
\refs{\wadia,\desing,\deboeretal}. It turns out that the conical metric is not a good approximation to the long
distance behavior of  \internal\ since \internal\ contains terms that decay slowly at $ r \to \infty$. In fact
there is a massless field in $AdS$ with conformal weight two ($ \Delta = \bar \Delta =1$) which has a vev, this
implies that the metric \internal\ differs from the $AdS$ metric by terms of order $1/r^2$. That is precisely the
order of the difference between the conical metrics and the $AdS$ metric. This is discussed in more detail in
appendix B.

It is interesting to consider the limit of very large $m$ with
$\beta$ fixed. In that limit we can set $\alpha =1$ as long
as we are at a distance of order $R_{AdS}/\sqrt{ m}$ from the
line at $r=0, ~\theta = \pi/2$.
The limiting metric  is obtained from \internal\ by setting
$\alpha =1$.
In this limit the solution has singularity along the
circle $r=0,\theta={\pi\over 2}$ like the one present in the
Aichelburg-Sexl  metric \aich .  The metric
 coincides with the  solution in \lmRot , which was expected
to describe the metric of high momentum particles moving along a
maximum circle of $S^3$.

Another interesting limit that we can take is a ``plane wave''
limit where we  concentrate on distances which are small compared
to the $AdS$ radius. In this limit, and  in the region where
$\alpha =1$, the metric  is
\eqn\PPlimit{\eqalign{ ds^2&=
2dx^+dx^- -(s^2+u^2)(dx^+)^2+ds^2+du^2+u^2d{\tilde\psi}^2+
s^2d\chi^2\cr &+(\beta-1)\left[2dx^+dx^- -(s^2+u^2)(dx^+)^2-
{(dx^-)^2\over s^2+u^2}\right]\cr &+{\beta-1\over s^2+
u^2}\left[u^4d{\tilde\psi}^2- 2s^2u^2d\chi d{\tilde\psi}
+s^4d{\chi}^2\right] }} where $x^+ = t$, $x^- = \phi
R^2_{AdS}$ and ${\tilde\psi}=\psi+\chi$. This metric is singular at $s=u=0$,
but close to this
point it is also necessary to take into account the full form for
$\alpha$ which we give in appendix B. The final metric is
non-singular and is explicitly written in appendix B. We see that
the behavior near  $r=s=0$ is of the form we expect for a metric
which is carrying  momentum density $p_- \sim (\beta-1)$ in
the $\phi$ or $x^-$ direction.

An interesting  aspect of \PPlimit\
is that the metric does not asymptote to
the plane wave metric at large  $u^2 + s^2 $. This is due to the
fact that from the plane wave point of view we have constant $p_-$
{\it density} and therefore infinite total $p_-$.
We show in section 4 that solutions  with excitations localized
in the $x^-$ direction which carry finite total $p_-$ are indeed
asymptotic to the standard plane wave.

An interesting question we would like to understand is the behavior of the energy gap in these geometries. If we
concentrate on excitations that are $\phi$ and $\psi$ independent then the energy gap can be computed easily in
the case of $m=1$ where we obtain (see appendix B) \eqn\gapmone{ \omega_0 =2\sqrt{\beta} } for the energy of the
lowest energy excitation. The energy gap for large $m$ is harder to estimate but we prove in appendix B that it is
always lower than \gapmone .

We see that as we increase the energy of the solution (by decreasing
$\beta$, see \nsangmom ) the
redshift factor at the origin decreases, so that a clock
runs more slowly there and also the energy gap to the next
excitation is very low.

\subsec{Remarks on the CFT description}

 A semi-quantitative explanation of this
fact was given in \lmBH \lmRot . The idea is that these chiral
primaries will involve multiply wound strings. The energy gap for
exciting such states becomes smaller as $1/w$ where $w$ is the winding.
This decrease of the energy gap can be seen
 even  when the CFT is at its orbifold point,
 where the
theory becomes a free CFT whose target space is $Sym(T^4)^{n_1 n_5}$
(for a full discussion of this system see \review ).
The NS ground state is in the untwisted sector and can be interpreted
as consisting  of $n_1 n_5$ singly wound strings.   The energy
gap to the next BPS excitation is of order one (we normalized the
circle of the CFT to have radius one).
High angular momentum single particle
chiral primaries  involve strings that are multiply wound.
Non-BPS excitations on a string of winding number $w$ go as
$2/w$.
In order to obtain a more precise match with the particular energy
gaps we obtained above it seems that we need to go away from the
orbifold point otherwise we can easily run into contradictions.
For example, let us
 consider the chiral primaries with $ ( J_L , J_R)
= (1/2,1/2)$ that come from the internal torus. These would
naively correspond, in the free orbifold picture, to
states in the untwisted sector (singly wound strings) where
we have some excitations on some of the $n_1n_5$ singly wound
 strings. More precisely,
each single particle excitation corresponds to exciting one string by adding a left and right moving fermion in
the lowest state. Since all strings are singly wound we get an energy gap of order one at the orbifold point
independently of $\beta$. On the other hand, the gravity description of such states is given by \internal\ with
$m=1$, for which the energy gap is \gapmone . Clearly we need to take into account that the supergravity picture
is valid only when we get away from the orbifold point in the CFT, which blurs the distinction between singly
wound and multiply wound strings. It would be highly desirable to understand better this effect from the CFT point
of view. In \lmBH\ some agreement was found with this naive picture, since only very special geometries and chiral
primaries were used.


\newsec{Solutions with plane wave asymptotics}

 {}From the general solutions in \DoDfChiral\ it is also possible to
obtain a general family of solutions with plane wave asymptotic boundary
conditions. The final prescription is that in order to obtain such
solutions we should drop the ones in \HarmFunc\ and consider a
profile ${\bf F}(v)$ which is a straight line in $R^4$  with small wiggles
in the various $R^4$ and $T^4$  coordinates.
 This is rather analogous to our previous
discussion where we took a profile that was a circle with some wiggles.

First we discuss how to obtain this as a limit of the general metric
\DoDfChiral\ corresponding to a circular profile with some small
oscillations. It is a limit where we zoom in into a small section of the
circle where this section looks like a straight line with some
oscillations.
More precisely, we rescale the coordinates and the profile:
\eqn\PPscale{\eqalign{
&x_1=a+{x'_1\over R},\quad x_2={x'_2\over R},\quad x_3={x'_3\over R},\quad
x_4={x'_4\over R},
\quad t=t'R,\quad y=\chi R,\quad
R={{\hat R}\over \eps},\cr
&F_1=a\cos\omega v+{{\hat F}_1({{v}R})\over R},
\quad F_2=a\sin\omega v+{{\hat F}_2({vR})\over R},\quad
F_3={{\hat F}_3({vR})\over R},\quad F_4={{\hat F}_4({vR})\over R}
}}
and we define the new rescaled functions
$$
{\hat f}_1=\eps^2 f_1,\quad {\hat f}_5={\eps^2{\hat R}^4} f_5,\quad
{\hat A}_i={1\over R^2} A_i,\quad {\hat B}_i={1\over R^2} B_i,
$$
Then we take a limit $\eps\rightarrow 0$ while the new coordinates and
${\hat R}$ remain fixed.

Defining the new parameter $\sigma = a \omega{\hat R}^2$ and dropping the primes in the new coordinates we find
\eqn\PPffive{\eqalign{ &{\hat f}_5={1\over 2\pi}\int_{-\infty}^\infty {dv'\over (x_2-{\hat F}_2- \sigma
v')^2+(x_\perp-{\hat F}_\perp)^2}\cr &{\hat f}_1={1\over 2\pi}\int_{-\infty}^\infty {\left[ (\sigma +{\dot {\hat
F}}_2)^2+|{\dot {\hat F}}_\perp|^2\right]dv'\over (x_2-{\hat F}_2- \sigma v')^2+(x_\perp-{\hat F}_\perp)^2}\cr
&{\hat A}_2=-{1\over 2\pi}\int_{-\infty}^\infty { (\sigma +{\dot {\hat F}}_2)dv'\over (x_2-{\hat F}_2- \sigma
v')^2+(x_\perp-{\hat F}_\perp)^2}, \cr &{\hat A}_\perp=-{1\over 2\pi}\int_{-\infty}^\infty { {\dot {\hat F}}_\perp
dv'\over (x_2-{\hat F}_2- \sigma v')^2+(x_\perp-{\hat F}_\perp)^2} }} Here we introduced a new integration
parameter $v'=vR$. Note that in terms of this parameter the argument of ${\hat F}_i$ is $\eps$--independent. In
the derivation of \PPffive~ we also used the relation \LDoDf
$$
{Q\over L}={R\over 2\pi}
$$
Note that the functions $\hat F^i(v')$ are not required to be
periodic, they are arbitrary functions of $v'$.
In terms of the functions \PPffive~ the metric becomes
\eqn\PPmetTwo{
ds^2 ={1\over \sqrt{{\hat f}_1{\hat f}_5}}\left[-(dt-{\hat A}_idx^i)^2+
(dy+{\hat B}_idx^i)^2\right]+\sqrt{{\hat f}_1{\hat f}_5}dx^idx^i
}
where ${\hat B}$ is defined by
$$
d{\hat B}=-^*d{\hat A}
$$

Let us examine the asymptotic behavior of this metric at large
$x_\perp$. In this limit ($x_\perp\gg |{\hat F}|$) we get the
following leading contributions to the harmonic
functions\foot{Here we assumed that $\langle {\dot{\hat
F}}_i\rangle=0$. If this condition is not true, then string moves
along the direction $\langle {\dot{\hat F}}_i\rangle$ and we can
account for this motion by redefining coordinate $x_2$.}
\eqn\PPlimitField{\eqalign{ &{\hat f}_5\approx{1\over 2 \sigma
x_\perp},\quad {\hat f}_1\approx{1\over\beta}{\sigma \over
2x_\perp},\quad {\tilde A}_2\approx-{1\over 2x_\perp},\quad {\hat
B}_\psi\approx-{(\cos\theta-1)\over 2}, }} all other components
of the gauge fields ${\hat A}$ and ${\hat B}$ are subleading. In
\PPlimitField~ we introduced a parameter $\beta\le 1$:
\eqn\PPgamma{ \beta \equiv \left(1+{1\over \sigma^2}\langle({\dot
{\hat F}}_2)^2+ |{\dot {\hat F}}_\perp|^2\rangle\right)^{-1} }
Substituting the expressions \PPlimitField~ in \PPmetTwo~ and
introducing the new coordinates: \eqn\coordredef{\eqalign{ x^+=&
{t} ~,\quad x^-= {x_2 \over \sqrt{\beta}},\quad
{\tilde\psi}=\psi+\chi \cr u=& \beta^{-1/4} \sqrt{2x_\perp
}\sin{\theta\over 2},\quad s=\beta^{-1/4} \sqrt{2x_\perp
}\cos{\theta\over 2}, }} we find \eqn\PPmetFive{\eqalign{
ds^2&=-\beta \left[2dx^+dx^-+(u^2+s^2) (dx^+)^2 \right] + (1 -
\beta) { (dx^-)^2 \over u^2+s^2} \cr &+du^2+ds^2+ {u^2s^2\over
u^2+s^2}(d{\tilde\psi}-d\chi)^2 +{\beta\over u^2+s^2}(s^2
d\chi+u^2d{\tilde\psi})^2 }} which is indeed the same as
\PPlimit\ after some simple changes of signs.
 This is the general behavior of the metric for
a configurations with uniform momentum density in the $x^-$ direction. If the excitation is localized in the $x^-$
direction, as we expect it to be for a finite $p_-$ wavepacket, then the profile for the corresponding vibration
will be such that $|{\dot{\hat F}}|$ differs significantly from zero only in a finite range of $v'$. Then  the
averages entering the expression \PPgamma~ vanish and $\beta=1$. So in this case the metric asymptotically goes to
the usual plane wave.\foot{ Note, in particular, that the coefficient of the term that goes as $(dx^-)^2/(s^2 +
u^2)$ goes to zero, while in flat space it goes over to some function of $x^-$. The difference is due to the rapid
decay in the transverse coordinates of wavefunctions with fixed $p_-$. }

So far we have been looking at  profiles which oscillate only in the
noncompact directions. For the case of oscillations
on the torus the six dimensional Einstein metric is still given by \PPmetTwo, but the function ${\hat f}_1$ should
be replaced by ${\bar f}_1$ \eqn\PPtor{\eqalign{ {\bar f}_1&= {1\over 2\pi}\int_{-\infty}^\infty {\left[
(\sigma+{\dot {\hat F}}_2)^2+|{\dot {\hat F}}_\perp|^2+ |{\dot {\hat {\cal F}}}|^2 \right]dv'\over (x_2-{\hat
F}_2-\sigma v')^2+(x_\perp-{\hat F}_\perp)^2} -{\hat f}_5^{-1}{\hat{\cal A}}_a{\hat{\cal A}}_a, \cr {\hat{\cal
A}}_a&=-{1\over 2\pi}\int_{-\infty}^\infty { {\dot {\hat{\cal F}}}_a dv'\over (x_2-{\hat F}_2-\sigma
v')^2+(x_\perp-{\hat F}_\perp)^2} }} The large $x_\perp$ limit of the resulting solution still has the form
\PPmetFive, but $\beta$ is now defined by
$$
\beta \equiv \left(1+{1\over \sigma^2}\langle({\dot {\hat F}}_2)^2+
|{\dot {\hat F}}_\perp|^2+|{\dot {\hat {\cal F}}}|^2\rangle\right)^{-1}
$$

\subsec{The solution for a localized excitation}

In order to understand the asymptotic behavior of the solution when the excitation is localized in $x^-$, we write
down the explicit form for one such solution with a very simple profile. We pick a profile with a perturbation
 only in the torus direction.
$$
{\hat{\cal F}}_1=\left\{\matrix{&0,\quad &|v'|>v_0>0\cr
&b(v_0-|v'|),\quad &|v'|<v_0\cr}\right.
$$
with all other components of ${\hat{\cal F}}_a$ and
${\hat F}_i$ equal to
zero.
This profile gives following harmonic functions
\eqn\PPExOne{\eqalign{
&{\hat f}_5={1\over 2 \sigma x_\perp},\quad
{\tilde A}_2=-{1\over 2x_\perp},\quad
{\hat B}_\psi=-{(\cos\theta-1)\over 2},\cr
&{\cal A}_1={b\over 2\pi \sigma  x_\perp}
\arctan\left\{{2(\sigma v_0)^2 x_\perp x^-\over
[x_\perp^2+(x^-)^2]^2+(\sigma v_0)^2(x_\perp^2-(x^-)^2)}
\right\}
\cr
&{\bar f}_1={\sigma \over 2x_\perp}+
{b^2\over 2\pi \sigma x_\perp}
\arctan \left({2\sigma v_0x_\perp\over x_\perp^2+(x^-)^2-(\sigma v_0)^2}
\right)
-{\hat f}_5^{-1}{\hat{\cal A}}_a{\hat{\cal A}}_a,
}}
In particular for $\sigma
 v_0\ll x_\perp$ and arbitrary value of $x^-$ we get
$$
{\bar f}_1={\sigma \over 2x_\perp}+
{b^2\over 2\pi \sigma  x_\perp}
{\sigma  v_0\over x_\perp}
\left\{{2 x_\perp^2\over (x^-)^2+x_\perp^2}+
O\left({\sigma  v_0\over x_\perp}\right)
\right\}
$$
Thus in the leading order at large $x_\perp$ we get a usual plane wave \citePP , i.e. the metric \PPmetFive~ with
$\beta =1$.


\newsec{The supertubes in different dimensions}

The previous analysis of the D1-D5 system is special because the
configuration blows up to a Kaluza-Klein monopole  and leads to a
non-singular situation. This will be the case also for all
configurations with two charges that result from doing U-duality
on the $T^4$. Of course, if we did T-duality on the $S^1$ where
the D1 and D5 are wrapped we would get a singular metric since
the KK monopole would become an NS 5 brane. The fact that the
metric is non-singular is related to the fact that the theory has
a non-zero energy gap for generic non-BPS excitations around the
state with maximal angular momentum. This energy gap is also
non-zero for other two charge systems in different number of
dimensions. So we considered similar supergravity solutions in
different dimensions but  we found that they were all singular.
In this section we summarize this discussion. For simplicity we
have concentrated on the solutions with maximal angular momentum
in a given two plane of the non-compact transverse directions.
Another question we consider is the following. We take the large
radius limit of the ring and then look at the resulting geometry.
It turns out that the following two  limits do not commute,
 the near
ring limit for fixed ring radius and the large radius first
and then the small distance limit. The physical reason why they
do not commute is that when one is approaching the ring one is
exploring the IR region of the field theory living on the branes
so that one is sensitive to the long distance geometry of the branes.

\subsec{Solutions in different dimensions}

In order to analyze such systems, we start with the F1-P1 system
in $R^{1,d}\times S^1\times  T^{8-d}$
(with the appropriate powers in the
harmonic functions) , integrate the string sources along a ring
(as explained in \FP ), and then perform some dualities on it to
get the desired system.
We can write the solutions in different U-dual frames.
We choose to work with the D0-F1
system blowing up to a D2 - the supertube of \supertubes .

We start then with the 1/4 supersymmetric supergravity solution
describing an oscillating string, wound around the $S^1$ and
 carrying right moving momentum
\MCP, \Harvey~ in $d$ non-compact  transverse  dimensions
\eqn\helstsol{\eqalign{ ds^2 &= -e^{2\Phi} dudv -
(e^{2\Phi}-1)\dot{F}^2 dv^2 +2(e^{2\Phi}-1)\dot{\bf{F}} \cdot
d{\bf x}dv + d{\bf x}^2_{d} +d{\bf z}^2_{8-d} \cr &\equiv
H({\bf x},v)(-dudv+K({\bf x},v)dv^2+2A_i({\bf x},v)dx^idv)+
d{\bf x}^2_{d} +d{\bf z}^2_{8-d}\cr B_{uv} &= {1\over 2}
(e^{2\Phi}-1) ~~~~~;~~~~~ B_{vi} = -\dot{F}_i(e^{2\Phi}-1) =HA_i
\cr e^{-2\Phi} &= 1+{Q \over {|{\bf x}-{{\bf F}(v)}|^{d-2} }} \cr }}
where the light cone coordinates are $u,v= t\pm y$ with  $y\sim
y+R_y$ and ${\bf x}$ are $d$ noncompact directions and ${\bf z}$
parameterize  a $T^{8-d}$. ${\bf F}(v)$ is a $d$-dimensional
vector
describing the location of the string. 

Taking the ring profile: $(F^\alpha_1+iF^\alpha_2)(v)=ae^{i(\omega
v +\alpha)}~,~F_3^\alpha=F_4^\alpha=0$, and integrating the harmonic functions along
$\alpha$, we get functions describing oscillating strings
uniformly distributed along a ring:
\eqn\smeared{\eqalign{
\langle H^{-1}({\bf x})\rangle  &= 1+{Q \over {2\pi}}\int_0^{2\pi}
{{d\alpha}\over {|{\bf x}-{\bf F}^{\alpha}|^{d-2}}} = 1+{Q\over
\sigma^{d-2}}I_1^{(d-2)}(-{2as\over \sigma^2})
\cr \langle K({\bf x})\rangle  &= {Q \over
{2\pi}}\int_0^{2\pi} d\alpha {|\partial_v{\bf F}^\alpha|^2d\alpha
\over {|{\bf x}-{\bf F}^{\alpha}|^{d-2}}} =
a^2\omega^2(\langle H^{-1}\rangle -1) = {a^2\omega^2 Q\over
\sigma^{d-2}}I_1^{(d-2)}(-{2as\over \sigma^2})\cr
\langle A_{\phi}({\bf x})\rangle  &=
as\omega {Q\over {2\pi}}\int_0^{2\pi} {\partial_vF_{\phi}
d\alpha\over {|{\bf x}-{\bf F}^{\alpha}|^{d-2}}} = {a\omega
Qs\over \sigma^{d-2}}I_2^{(d-2)}(-{2as\over \sigma^2}) }}
where $s^2 \equiv
x_1^2+x_2^2$ is the radial coordinate in the ring plane, $w^2
\equiv x_3^2 + \cdots+x_{d}^2$ is the perpendicular distance
from the ring plane, $\sigma^2\equiv a^2+s^2+w^2$, and where we
defined the integrals: \eqn\defis{ I_1^{(n)}(k) \equiv {1\over
{2\pi}}\int_0^{2\pi} {{d\alpha}\over{(1+k\cos\alpha)^{n/2}}}
~~~~;~~~~ I_2^{(n)}(k) = {1\over {2\pi}} \int_0^{2\pi}
{{\cos\alpha d\alpha}\over{(1+k\cos\alpha)^{n/2}}} } for integer
$n$'s. The integrals above can be easily evaluated, and appear in
appendix D. For odd $n$ they involve elliptic functions.

Now, if we perform S-duality and then $T_y$ duality on \helstsol , we get the $F1D0\to D2$ supertube solution for
any dimension $d$, with $H,K,A_{\phi},I_1,I_2$ as in \smeared , \defis \eqn\DzeroFone{\eqalign{ ds^2 &=  {1 \over
\sqrt{H^{-1}}(1+K)}\Big[ -dt^2+2A_\phi dt d\phi +
 H^{-1} dy^2+\cr &+s^2({a\omega Q\over
\sigma^{d-2}})^2\{((I_1^{(d-2)})^2-(I_2^{(d-2)})^2)+{1+a^2\omega^2\over
a^2\omega^2 Q}\sigma^{d-2}I_1+({\sigma^{d-2}\over a\omega
Q})^2\}d\phi^2 \Big] +\cr &+ \sqrt{H^{-1}}\Big(
ds^2+d{\bf w}^2_{d-2}+d{\bf z}^2_{8-d} \Big) \cr e^{-2\Phi} &=
H^{3/2}(1+K) \cr B_2 &= -{K \over {1+K}}dt\wedge dy- {A_\phi \over
{1+K}}d\phi\wedge dy \cr C_1 &= -(H-1) dt+ HA_\phi d\phi ~~~~;~~~~
C_3 = -{{A_\phi}\over {1+K}} dt\wedge d\phi\wedge dy \cr }} where
in the second line we wrote $g_{\phi\phi}= {1 \over
\sqrt{H^{-1}}(1+K)}(s^2H^{-1}(1+K)-A_\phi^2)$ in terms of the
integrals $I_1,I_2$. \foot{ Note that $I_1^{(d-2)},I_2^{(d-2)}$
in \DzeroFone~ are evaluated at $-{2as\over \sigma^2}$.}

Analyzing the behavior of the different functions in the metric
above, one finds that these solutions are everywhere regular,
except maybe on a ring of radius $a$ in the $x$-space ($w=0 ,
s=a$). In appendices C and D   we show that these solutions are
indeed singular on the ring in all dimensions except the for the
D1-D5 system described above. We did not find any U-dual frame
where the solutions were regular, except for $d=4$ which is
U-dual to the D1-D5 system \foot{Recently the $d=3$ case was
analyzed in \Lugo.}.

\subsec{The large ring limit}

In this subsection we take the limit where the radius of the ring
becomes very large. This can be achieved by taking very large values
of the charges.  From the formula of the ring radius
\MaxA\ we see that in this limit we expect to have a finite
energy density per unit length along the ring.
In this limit the ring becomes a straight line.
We want to find the metric near the ring in this case.
This metric has the form of the metric of a brane with some fluxes
on it. In the D1-D5 it will be a KK monopole with some fluxes on it.
These fluxes have a special value such that the supersymmetry that
is preserved is independent of the orientation of the brane.
Below we explain this in detail.

We can try take the limit where $a\to\infty$ in \DzeroFone , and see
if the solutions we obtain really describe a flat $D2$ with
fluxes.

 {} From a worldvolume analysis \supertubes~ one finds that the radius
of the supertube scales with the D0,F1 charges as $a\sim \sqrt{Q_0Q_s}$ where the two charges in our notations are
$Q , a^2\omega^2 Q$. Keeping the ratio of the charges fixed, the scaling is \eqn\scaling{ a \sim Q \to \infty
~~~~;~~~~ \delta \equiv a\omega ~fixed }

Taking this limit for a fixed $\rho$ in \DzeroFone~ and defining
the coordinate $x_\|\equiv a\phi$ gives the following metrics and
fields for $d\geq 4$: \eqn\generald{\eqalign{ ds^2 &=
{[1+{\tilde{q}\over\rho^{d-3}}]^{1/2}\over
[1+\delta^2{\tilde{q}\over\rho^{d-3}}]}\{ dy^2 + {1+ {\tilde{q}
\over \rho^{d-3}}(1+\delta^2)\over 1+{\tilde{q} \over
\rho^{d-3}}}dx_\|^2+{2\delta{\tilde{q} \over \rho^{d-3}}\over
1+{\tilde{q} \over \rho^{d-3}}}dx_\|dt-{dt^2\over
{1+{\tilde{q}\over \rho^{d-3}}}}\} +\cr &+
[1+{\tilde{q}\over\rho^{d-3}}]^{1/2}[d\rho^2+\rho^2d\Omega_{d-2}^2+
dz_{18-d}^2] \cr B_2 &= -{\delta{\tilde{q}\over \rho^{d-3}}\over
1+\delta^2{\tilde{q}\over \rho^{d-3}}}(\delta dt\wedge
dy+dx_\|\wedge dy )\cr C_1 &= { {\tilde{q}\over \rho^{d-3}}\over
1+{\tilde{q}\over \rho^{d-3}} } (dt+\delta dx_\|) ~~~~~;~~~~~  C_3
= -{\delta{\tilde{q}\over \rho^{d-3}}\over
1+\delta^2{\tilde{q}\over \rho^{d-3}}}dt\wedge dx_\|\wedge dy \cr
e^{-2\phi} &= (1+{\tilde{q}\over
\rho^{d-3}})^{-3/2}(1+\delta^2{\tilde{q}\over \rho^{d-3}})}}
where the effective charge $\tilde{q}$ is given by $$ \tilde{q} =
{Q\over a}\cdot \lim_{\rho\to 0}[({\rho\over
a})^{d-3}I_1^{(d-2)}(1-{{\rho^2\over 2a^2}\over 1+{\rho\over
a}\sin\Theta +{\rho^2\over 2a^2}})] $$ which for the different
dimensions is : \eqn\effq{\matrix{ &d=&4 &5&6&7 &8 \cr
&\tilde{q}=& {Q\over 2a} &{Q\over \pi a} &{Q\over 4a} &{2Q\over
3\pi a} &{3Q\over 16a} \cr}} For $d=3$ we get a logarithmic
singularity .

We would like to compare 
\generald~ with the metric and fields describing a D2-brane with
F1 and D0 fluxes on a $T^{8-d}$. These can be generated by
starting with the supergravity solution of a $D2$ in the
$\tilde{t},\tilde{y},\tilde{x_p}$ directions \foot{We choose a
gauge where the Ramond-Ramond Gauge field vanishes at spatial
infinity.}. Then T-dualizing in $\tilde{x_p}$ to obtain a D1 in
the $\tilde{y},\tilde{t}$ directions, smeared on the
$\tilde{x}_p$ direction (with a harmonic function $f=1+{q\over
r^{d-3}}$) .Then making a boost and a rotation with parameters
$\alpha,\theta$ mixing $\tilde{t},\tilde{y},\tilde{x}_p$ to give
$t,y,x_p$  \foot{so that
$$\eqalignno{
\tilde{t}&=\cosh\alpha t +\sinh \alpha(\cos\theta x_p+\sin\theta y) \cr
\tilde{x}_p&= \cosh\alpha(\cos\theta x_p+\sin\theta
y)+\sinh\alpha t \cr \tilde{y} &=(\cos\theta y-\sin\theta
x_p)
}$$ }, and finally making a T-duality in the $y$-direction \foot{this procedure was explained for example in
\MalRusso} .

This gives the following metric and gauge fields : \eqn\fluxdtwo{\eqalign{ ds^2
&=f^{-1/2}[-(1-h^{-1}{q\sinh^2\alpha\over r^{d-3}})dt^2+(1+h^{-1}{q\cosh^2\alpha\cos^2\theta\over r^{d-3}})dx_p^2+
\cr &~~~~~~~~~~+ 2h^{-1}{q\sinh\alpha\cosh\alpha\cos\theta \over r^{d-3}}dtdx_p +
fh^{-1}dy^2]+f^{1/2}[dr^2+r^2d\Omega_{d-2}^2+dz^2_{8-d}] \cr B_2 &=  -h^{-1}{q\sin\theta\cosh\alpha\over
r^{d-3}}[\sinh\alpha dy\wedge dt+\cosh\alpha\cos\theta dy\wedge dx_p] \cr C_1 &= (f^{-1}-1)[\cos\theta\cosh\alpha
dt+\sinh\alpha dx_p] ~~~;~~~ C_3 =h^{-1}{q\sin\theta\cosh\alpha\over r^{d-3}} dt\wedge dx_p\wedge dy\cr  e^{2\phi}
&=g^2f^{3/2}h^{-1} \cr f &\equiv 1+{q\over r^{d-3}} ~~~~;~~~~ h\equiv 1+{q\cosh^2\alpha\sin^2\theta\over r^{d-3}}
}}

Comparing \fluxdtwo~ to \generald~ we find exact agreement if we choose $\sinh\alpha=\tan\theta=\delta$. All of
the solutions \fluxdtwo~ are 1/2 supersymmetric as they are dual to a D2. However only the subfamily of such
solutions with $\sinh\alpha=\tan\theta$ would continue being supersymmetric (with 1/4 supersymmetry) if we start
curving the brane, taking the direction $x_p$ and putting it on some closed curve, e.g. the ring. (The exact
supersymmetries that this curvely shaped D2 with fluxes preserves can be found doing a worldvolume analysis , as
done in \supertubes~ , or as done for the $D2\overline{D2}$ system in \DDbar ). Under a T-duality in the $S^1$
circle this system becomes a D1 brane that winds along the $S^1$, and moves along the $S^1$  as it stretches in
the $x_p$ direction. The velocity is such that a brane that is stretched in the opposite direction along $x_p$ but
with the same winding and velocity intersects the original brane at a point that moves with the speed of light
\myerslight . These configurations preserve 1/4 of the supersymmetries. These configurations are intimately
related to the oscillating strings we started with. In fact strings carrying oscillations only in one direction
will intersect with each other at points that move at the speed of light.

In the $d=4$ case we can make a U-duality to the D1-D5 system so that the  large ring radius \generald\  becomes a
straight KK monopole carrying D1 and D5 fluxes
\eqn\lim{\eqalign{ ds^2 &= [1+{q_1\over\rho}]^{-1/2} [1+{q_5\over
\rho}]^{-1/2} [-(dt-{\sqrt{q_1q_5}\over \rho}dx_\|)^2+(dy-{\sqrt{q_1q_5}} (1-\cos\Theta)d\psi)^2] \cr &+
[1+{q_1\over\rho}]^{1/2} [1+{q_5\over \rho}]^{1/2}[d\rho^2+\rho^2d\Theta^2+\rho^2\sin^2\Theta
d\psi^2+dx_\|^2]+\sqrt{\rho + q_1\over \rho + q_5 } dz_{(4)}^2 \cr e^{2\phi} &= { \rho + q_1 \over \rho + q_5 }
\cr C_2 &= -{ q_1\over \rho + q_1 }(dt+\sqrt{q_5\over q_1}dx_\|)\wedge (dy +\sqrt{q_5\over
q_1}\rho(1-\cos\Theta)d\psi) }} where we have defined the charge densities $q_i = Q_i/(2 a) $ which are finite in
the limit. This metric is non-singular if $R_y = 2 \sqrt{q_1 q_5} $. This is a condition on the fluxes for a given
radius $R_y$. If we U-dualize \fluxdtwo\ we can get solutions which represent KK monopole with arbitrary values of
the fluxes that are 1/2 BPS. What is special about the fluxes in \lim\ is that we can reverse the KK monopole
charge, keeping the same values for $q_1, q_5$ so that the configuration with KK and anti-KK charges
 still preserve 1/4 of the  supersymmetries. As shown in
\myerslight\ this configuration is U-dual to configurations
with intersecting D-branes where the intersection point moves
at the speed of light (see also \bachas).

Note that in the limit that we drop the 1 in the harmonic functions
that appear in \lim\ we obtain a plane wave in six dimensions.
We can get this as a limit where we scale the charges to
infinity and the rest of the coordinates appropriately.
The geometry \lim\ thus provides us with a spacetime which
is asymptotically flat and that looks like a plane wave in a  suitable
near horizon limit.

 \centerline{\bf Acknowledgements}

We would like to thank Samir Mathur and Sameer Murthy
for useful discussions.
This work was supported in part by DOE grant DE-FG02-90ER40542
and
NSF grant PHY--0070928.

\appendix{A}{ Topology of the solutions}

If we have a single ring profile, such as the one in
\eqa\ then the harmonic functions \HarmFunc\
can be found explicitly and read
\eqn\ringharm{\eqalign{
f_5-1 =& Q h^{-1}~,~~~~~ f_1-1 = a^2 \omega^2 Q h^{-1}
\cr
h^2 =& [ (s+a)^2 + w^2][(s-a)^2 + w^2 ]
\cr
A_\phi = & 2 a^2 \omega  Q s^2 { 1 \over h ( h + s^2 + a^2 + w^2 )}
}}
where $s^2 = x_1^2 + x_2^2$ and $w^2= x_3^2 + x_4^2$.

In order to understand more clearly the topology of the metric
it is convenient to write the metric in other coordinates such
that the metric reads
 \eqn\DoneDfive{\eqalign{ ds^2 &= {1 \over
\sqrt{f_1f_5}}[-(dt- {{a\sqrt{Q_1Q_5}}\over
r^2+a^2\cos^2\theta}\sin^2\theta d\phi)^2+
(dy+{{a\sqrt{Q_1Q_5}}\over r^2+a^2\cos^2\theta}\cos^2\theta
d\psi)^2] +\cr &+ {\sqrt{f_1f_5}}
[(r^2+a^2\cos^2\theta)({{dr^2}\over {r^2+a^2}}+d\theta^2)+
r^2\cos^2\theta d\psi^2 + (r^2+a^2)\sin^2\theta d\phi^2]\cr &+
\sqrt{{f_1 \over f_5}} dz_adz_a \cr e^{2\phi} &= {f_1\over f_5},
\cr C_{(2)} &= (1-{1\over f_5})(dt- \sqrt{Q_1\over
Q_5}a\sin^2\theta d\phi)\wedge (dy-\sqrt{Q_1\over Q_5}
a\cos^2\theta d\psi)\cr &+Q_1\cos^2\theta d\phi\wedge d\psi \cr
f_{1,5} &= 1 + { Q_{1,5} \over r^2 + a^2 \cos^2\theta } }} This
form of the metric arises naturally if we view the solution as a
limit of the general five dimensional black hole solutions in
\cvetic .  The explicit coordinate change from the coordinates
$w,s$ in \ringharm\  to the ones in \DoneDfive\ is
\eqn\coordchange{ s^2=(r^2 + a^2 )\sin^2 \theta~,~~~~~ w=r \cos\theta }
and $\phi$ and $\psi$ are again the phases of $x_1+
i x_2$ and $x_3 + i x_4$ respectively. Here there is a potential
singularity when $r=0$ and $\theta = \pi/2 $. We can rewrite the
metric \DoneDfive~ in a form where its singularity structure is
more transparent \eqn\metrict{\eqalign{ ds^2 = & \sqrt{f_1
f_5}(r^2+a^2\cos^2\theta) [  d\theta^2 + h \cos^2 \theta ( d\psi
+ { a \sqrt{Q_1 Q_5} \over f_1 f_5 h(r^2+a^2\cos^2\theta)^2}
dy)^2 + \cr &~~~~~~+ \tilde h\sin^2 \theta  (d\phi + { a
\sqrt{Q_1 Q_5 } \over f_1 f_5 \tilde h (r^2+a^2\cos^2\theta)^2} d
t)^2 ] + \cr & + \sqrt{f_1 f_5 } (r^2+a^2\cos^2\theta)[ { r^2
\over g } dy^2 + { dr^2 \over r^2 + a^2 } ] - { 1 \over \sqrt{f_1
f_5} } ( 1 + { \sin^2 \theta a^2 Q_1Q_5 \over
 f_1 f_5 \tilde h (r^2+a^2\cos^2\theta)^3}) dt^2
}}

where the functions $h, \tilde h, g$ are
\eqn\funct{\eqalign{
g = &  Q_1Q_5 + (Q_1 + Q_5) r^2 + (r^2+a^2\cos^2\theta)r^2
\cr
h = & { g \over f_1 f_5(r^2+a^2\cos^2\theta)^2}
\cr
\tilde h = & { Q_1Q_5 + (Q_1 + Q_5) (r^2+ a^2) +
(r^2+a^2\cos^2\theta) (r^2+a^2)
\over f_1 f_5(r^2+a^2\cos^2\theta)^2}
}}
The important properties of these functions are
$ g(r=0,\theta) = Q_1 Q_5 $, $h(r, \theta =\pi/2) =1$,
$\tilde h(r, \theta =0) =1$.
These properties, together with \MaxA , ensure that the
solution is nonsingular.
Note that after the coordinate redefinition $\tilde \psi
= \psi + y/R $ and $ \tilde \phi = \phi + t/R$  the metric
near $r\sim 0$ looks like that of a deformed $S^3$.

In order to study the topology of the solution we notice that the
time direction will just give a factor of $R$, so we drop it from
the discussion. The topology of a surface of large $r$ is that of
$S^1 \times S^3$. Near $r=0$ we see that the $S^1$ circle shrinks
to zero size while the sphere parameterized by $\tilde \psi ,
\theta, \tilde \phi$ does not. Topologically this is basically
the same as the sphere we had at infinity since the map
$(y,\psi,\theta,\phi) \to (y, \tilde \psi, \theta, \tilde \phi)$
can be continuously deformed to the identity. This means that the
final topology of the spatial region $r \leq r_0$ is that of a
$D^2 \times S^3$. So the $S^3$ is non-contractible.

It is interesting  to understand what   the deformed three sphere
that we have at $r=0$ looks like in the original ``flat''
coordinates $s,w$. From \coordchange\ we see that $r=0$ is the
disk spanned by $w=0$ and $s<a$. On top of this we have the $y$
circle. These together from a three sphere since the y circle
shrinks to zero at the boundary of the disk.
Note that in the decoupling limit, where $Q_i $ become very large
the functions in \funct\ become constant. Then the three sphere
parameterized by the coordinates $\tilde \psi, \theta, \tilde
\phi$ is a round three sphere. Away from the decoupling limit it
is not metrically a round three sphere.

\appendix{B}{Gravity duals of chiral primaries on the torus.}

In the previous sections we discussed the geometries corresponding
to chiral primaries associated with AdS$_3\times $S$^3$. Such
chiral primaries are universal and they do not depend on the
structure of the internal manifold M in
AdS$_3\times$S$^3\times$M. But there are also some chiral
primaries associated with the internal manifold, and in this
section we will discuss them for the simplest case where $M=T^4$.
We comment on the K3 case at the end.

To construct the geometries corresponding to such chiral primaries, we will follow the steps outlined in section
2. Namely we will start from the vibrating string, perform the dualities to relate it to the D1-D5 system, and
then perform  spectral flow to go to the NS sector. The only difference is that now we will allow the string to
vibrate not only in noncompact directions, but also on the torus. Since to go to the D1-D5 system we have to
perform dualities in the torus directions, the geometry of the vibrating string should be translation invariant in
these directions, and we can achieve this by ``smearing'' in the torus coordinates (just like we smeared the
profile on the $y$ direction by performing integration over $v$ in the string profiles \HarmFunc ).

Thus we start with the metric of a vibrating string, smear it over
the torus directions, and perform the following dualities
\eqn\dualMap{{ P(5) \choose F1(5)}\matrix{&S\cr &\rightarrow}
{P(5) \choose D1(5)} \matrix{&T6789\cr &\longrightarrow}
{P(5)\choose D5(56789)}\matrix{&S\cr &\rightarrow}{P(5)\choose
NS5(56789)}\matrix{&T5\cr &\rightarrow}{F1(5)\choose NS5(56789) }}
This way we get an ``F1-NS5'' of type IIA theory\foot{ A simple
further T-duality in one of the $T^4$ directions would give a
solution in IIB} \eqn\TorMetr{\eqalign{ ds^2 &={1\over {\tilde
f}_1}\left[-(dt-A_i
dx^i)^2+(dy+B_idx^i)^2\right]+f_5dx^idx^i+dz^adz^a\cr
e^{2\Phi}&={f_5\over {\tilde f}_1},\quad B_{ty}=-1+{1\over
{\tilde f}_1},\quad B_{ti}={B_i\over {\tilde f}_1},\cr
B_{yi}&={A_i\over {\tilde f}_1},\quad
B_{ij}=C_{ij}-{A_iB_j-A_jB_i\over {\tilde f}_1}\quad
C^{(1)}_a=f_5^{-1}{\cal A}_a,\cr
C^{(3)}_{abc}&=f_5^{-1}\epsilon_{abcd}{\cal A}_d,\quad
C^{(3)}_{iya}={A_i{\cal A}_a\over {\tilde f}_1f_5},\quad
C^{(3)}_{ita}={B_i{\cal A}_a\over {\tilde f}_1f_5},\cr
C^{(3)}_{ija}&={(A_iB_j-A_jB_i){\cal A}_a\over {\tilde
f}_1f_5},\quad C^{(3)}_{tya}={\cal A}_af_5^{-1}\left(-2+{1\over
{\tilde f}_1}\right),\cr C^{(5)}_{tyabc}&=-\epsilon_{abcd}{\cal
A}_df_5^{-1} \left[2-{1\over {\tilde f}_1f_5}\right],\quad
C^{(5)}_{iyabc}=-\epsilon_{abcd}{A_i{\cal A}_d \over {\tilde
f}_1f_5},\cr C^{(5)}_{tiabc}&=-\epsilon_{abcd}{B_i{\cal A}_d
\over {\tilde f}_1f_5},\quad
C^{(5)}_{ijabc}=\epsilon_{abcd}{(A_iB_j-A_jB_i){\cal A}_d \over
{\tilde f}_1f_5}
 }}
$$
{\tilde f}_1\equiv f_1-f_5^{-1}{\cal A}_a{\cal A}_a
$$
One can now perform additional T duality along
one of the torus directions followed by S duality, to get a D1-D5
system. We will do this step only with the metric.
But
in any case, if one wants to study the properties of six
dimensional Einstein metric, then one gets the same results
starting either from D1-D5 or F1-NS5.
The
functions in \TorMetr\
 are given by\foot{The simplest way to construct the harmonic
functions  is following. We can first decompactify torus directions and look at the
vibrating string in eight noncompact directions. Then we can smear over positions of the string in
$z_1,\dots z_4$ (which corresponds to integration over ${\bf z}$ in $f_5$), and in the end
compactify $z_1,\dots z_4$ on the torus.}
\eqn\eqss{\eqalign{ f_5 &=1+{Q\over L}\int\int_0^L{d{\bf
z}dv\over \left[({\bf x}-{{\bf F}})^2+({\bf z}-{\bf{\cal
F}})^2\right]^2}= 1+{Q\over L}\int_0^L{dv\over ({\bf x}-{{\bf
F}})^2},\cr f_1 =&1+ {Q\over L}\int_0^L{|\dot G|^2 dv\over ({\bf
x}-{{\bf F}})^2},\quad A_i=-{Q\over L}\int_0^L{{\dot F}_i dv\over
({\bf x}-{{\bf F}})^2},\quad {\cal A}_a=-{Q\over L}\int_0^L{{\dot {\cal
F}}_a dv\over ({\bf x}-{{\bf F}})^2}. }} Here we introduced an
eight dimensional vector ${\bf G}=(F_i,{\cal F}_a)$.

Note that in \eqss\ we have integrated over the position ${\bf z}$ of the string in the internal torus. This is
done to obtain a solution that is independent of the internal coordinates. This implies that the dependence on
${\cal F}_a$ disappears from $f_5$ in \eqss\ , but does not disappear from $f_1$ and ${\cal A}_a$.

Let us analyze the metric \TorMetr near the singularity.
Near the singularity we get
 \eqn\eqt{\eqalign{ f_5= &{Q\over
L}{\pi\over |{\dot{\bf F}}|x_\perp},\quad f_1={Q|{\dot{\bf
G}}|^2\over L} {\pi\over |{\dot{\bf F}}|x_\perp},\quad
f_1 -1 - f_5^{-1} {\cal A}_a{\cal A}_a={Q\over L}{\pi |\dot{\bf F}|\over x_\perp},\cr
A_i=&-{Q\over L}{\pi{\dot F}_i\over |{\dot F}|x_\perp},\quad
{\cal A}_a=-{Q\over L}{\pi{\dot{\cal F}}_a\over |{\dot F}|x_\perp},\cr
}}
The expressions for $f_5$, $A_i$ and $f_1 -1 - f_5^{-1}
{\cal A}_a {\cal A}_a$
do not depend on the profile in the internal
directions ${\bf{\cal F}}$, and thus the criteria for the absence
of the singularity is the same as in the case with no vibrations
on the torus, namely  the profile should not self intersect in the
$x_1,\dots x_4$ space and ${\dot{\bf F}}$ should never vanish.

In the case of type IIB string theory on $AdS_3 \times S^3 \times
K3$ there are also chiral primaries that are associated to
extra anti-self dual 3-form gauge fields in six dimensions
 that come from anti-self-dual
two forms on $K3$. Using heterotic/IIA duality it is very simple
to get these solutions too. We have to perform the chain of
dualities \eqn\dualMap{ { P(5) \choose F1(5)}\matrix{&het/IIA\cr
&\rightarrow} {P(5)\choose NS5(56789)} \matrix{&T5\cr
&\rightarrow}{F1(5)\choose NS5(56789)} } so that in the end we
get a solution of IIB on K3. In the heterotic theory the
fundamental string can oscillate in the $T^4$ directions as well
as in the 16 extra bosonic left moving directions on the
heterotic worldsheet. Solutions of this type were discussed in
\MCP \Harvey . It is in principle straightforward to perform the
duality transformations, but we leave that for the interested
reader.

\subsec{Example of the vibration on the torus.}

We consider the simplest example for the vibrations on the torus: \eqn\CCprof{ F_1=a\cos\omega v,\quad
F_2=a\sin\omega v,\quad {\cal F}_1=b\cos m\omega v,\quad {\cal F}_2=b\sin m\omega v, } all other components are
zero. The frequency $\omega$ is related to the radius $R$ of the $y$ direction by \eqa. As we already mentioned,
the expressions for $f_5$ and $A_i$ remain the same as they were for $b=0$, so to find the metric we only have to
evaluate ${\tilde f}_1=f_1-f_5^{-1}{\cal A}_a{\cal A}_a$. Substituting the profile \CCprof in \TorMetr, we find:
\eqn\ExTorOn{ f_1=1+{Q\over r^2+a^2\cos^2\theta}\omega^2(a^2+b^2m^2),\quad {\cal A}_1=-{Q\over 2\pi}\sin m\phi
I_m,\quad {\cal A}_1={Q\over 2\pi}\cos m\phi I_m, } where \eqn\ExTorInt{\eqalign{ I_m&\equiv b\omega
m\int_0^{2\pi} {d\alpha\cos m\alpha\over {r^2+a^2\sin^2\theta+a^2-2a\sqrt{r^2+a^2}\sin\theta\cos\alpha}}\cr &=
{2\pi bm\omega\over r^2+a^2\cos^2\theta} \left(-{a\sin\theta\over \sqrt{r^2+a^2}}\right)^m }} Then we find:
\eqn\ExTorTwo{ {\tilde f}_1=f_1-f_5^{-1}{\cal A}_a{\cal A}_a=1+{Q\omega^2\over r^2+a^2\cos^2\theta} \left[
(a^2+b^2m^2)-{Qb^2m^2\over Q+r^2+a^2\cos^2\theta} \left({a^2\sin^2\theta\over {r^2+a^2}}\right)^{m}\right], } and
the metric for the D1--D5 system becomes: \eqn\CircCircMet{\eqalign{ ds^2 &= {1 \over \sqrt{{\tilde
f}_1f_5}}[-(dt- {{a^2 R}\over r^2+ a^2\cos^2\theta}\sin^2\theta d\phi)^2+ (dy+{{a^2 R }\over
r^2+a^2\cos^2\theta}\cos^2\theta d\psi)^2] +\cr &+ {\sqrt{{\tilde f}_1f_5}} [(r^2+a^2\cos^2\theta)({{dr^2}\over
{r^2+a^2}}+d\theta^2)+ r^2\cos^2\theta d\psi^2 + (r^2+a^2)\sin^2\theta d\phi^2]\cr &+ \sqrt{{{\tilde f}_1 \over
f_5}} d{\bf z}^2 }} Note that the total fivebrane charge is $Q_5 = Q$ and the onebrane charge is given by $Q_1 =
Q\omega^2(a^2 + m^2 b^2)$ where $\omega$ is as in \eqa . In particular we have:
$$
R= \sqrt{Q_1Q_5\over a^2+m^2b^2}
$$

In order to obtain \internal\ we need to drop the 1 in the
harmonic functions in \eqss . This gives
\eqn\harmAS{
\tilde f_1 -1 = {\alpha Q_1\over r^2+a^2\cos^2\theta},\quad
f_5-1={Q_5\over r^2+a^2\cos^2\theta},\quad
A_\phi=\sqrt{\beta}{a\sqrt{Q_1Q_5}\sin^2\theta\over r^2+a^2\cos^2\theta}
}
with $\alpha$ and  $\beta$ defined by
\eqn\defgama{
\beta = { a^2 \over a^2 + m^2 b^2},\quad \alpha =  1- (1 - \beta)
 \left({\beta \sin^2\theta\over {r^2+\beta}}\right)^{m}
} Note that for large values of $m$ $\alpha$ is equal to one everywhere except for the small vicinity of the ring
$r=0,\theta={\pi\over 2}$, and in the limit $m\rightarrow\infty$ the harmonic functions \harmAS reduce to the ones
for the solution corresponding to a ring of rotating particles \refs{\lmRot,\lms}.

We define $\chi = y/R$ and rescale
\eqn\rescal{
t \to R t,\quad
r^2 \to r^2 (a^2 + b^2 m^2) = r^2 { Q_1 Q_5 \over R^2},
}
then \ExTorOn becomes:
\eqn\internapp{\eqalign{
ds^2\over \sqrt{Q_1 Q_5}&=(r^2+\beta \cos^2\theta)
{1 \over\sqrt{\alpha}}\left[-(dt-{\beta \sin^2\theta
d\phi\over r^2+\beta \cos^2\theta})^2+
(d\chi+{\beta \cos^2\theta d\psi\over r^2+\beta \cos^2\theta})^2\right]\cr
&+{\sqrt{\alpha}dr^2\over r^2+\beta }+\sqrt{\alpha}d\theta^2+
{\sqrt{\alpha}\over r^2+\beta \cos^2\theta}(r^2\cos^2\theta d\psi^2+
(r^2+\beta)\sin^2\theta d\phi^2)
}}

Let us look at the limit $m\rightarrow\infty$ (which corresponds to $\alpha=1$) and compare the above metric with
the metric of the conical defect. To do this it is convenient to rewrite \internapp~ for $\alpha=1$ as
\eqn\ntsix{\eqalign{ ds^2\over \sqrt{{Q}_1 Q_5}&= -\left(r^2+{\beta-\beta^2\over
2}+\beta^2\right)dt^2+\left(r^2+{\beta-\beta^2\over 2}\right)d\chi^2 +{dr^2\over r^2+\beta}\cr
&+d\theta^2+\cos^2\theta(d\psi+\beta d\chi)^2 +\sin^2\theta(d\phi+\beta dt)^2\cr &+{(\beta-1)\beta\over
r^2+\beta\cos^2\theta}(\cos^4\theta d\psi^2- \sin^4\theta d\phi^2)+{\beta(1-\beta)\over 2}\cos
2\theta(-dt^2+d\chi^2) }} If we now introduce new coordinates:
$$
r'=r+{\beta-\beta^2\over 4r}(1-\cos 2\theta),\quad \theta'=\theta-
{\beta(\beta-1)\over 4r^2}\sin 2\theta
$$
then in the leading two orders at infinity the metric \ntsix~ becomes:
\eqn\ntseven{\eqalign{
ds^2\over \sqrt{{Q}_1 Q_5}&=
-\left({r'}^2+\beta^2\right)dt^2+{r'}^2d\chi^2
+{d{r'}^2\over {r'}^2+\beta^2}\cr
&+d{\theta'}^2+\cos^2\theta'(d\psi+\beta d\chi)^2
+\sin^2\theta'(d\phi+\beta dt)^2\cr
&+(\beta-1)\beta\cos 2\theta'\left[{1\over {r'}^2}(
d{\theta'}^2+\cos^2\theta'd\psi^2
+\sin^2\theta'd\phi^2
)+(dt^2-d\chi^2+{dr'^2\over r'^4})
\right]
}}
The first two lines give a metric of a conical defect, while the last line
gives a perturbation, which corresponds to an $AdS_3$ scalar with
angular momentum $l=2$. This mode is a mixture of an overall rescaling
of the sphere, AdS, and the three form field strengths. The fact that
the correction to the $AdS_3$ part of the metric in \ntseven\ is
not just an overall factor is due to the fact that these scalar
fluctuations also imply a change of the metric of the form
$\delta g_{\mu \nu} \sim \nabla_\mu \nabla_\nu \delta \phi$ where
$\delta \phi$ is the scalar fluctuation. More details and explicit
formulas can be found in \sezgincp .
Note that the terms in the last line of \ntseven\ are
of the same order as the terms of the $AdS_3$ part of the metric in
the first line. This implies that the conical defects are not
a good approximation to these metrics.

\subsec{ Plane wave limit of the solution}

In this subsection we take the plane wave limit of the solution
\internapp. Let us call $\sqrt{Q_1Q_5} = \epsilon^{-2}$, then
we define rescaled quantities by
\eqn\rescc{
t = x^+, ~~~~ \phi = \epsilon^2 x^- ~,~~~~ r = \epsilon
 \sqrt{\beta} s~,~~~~~
{\pi \over 2 } - \theta = \epsilon u ~,~~~~~ \tilde m = { m \over
\epsilon^2 } } In the $\epsilon \to 0$ limit we get the metric
\eqn\PPlimitfull{\eqalign{ ds^2&= \beta \alpha^{-1/2} [2dx^+dx^- -
(s^2+u^2)(dx^+)^2 ]+\alpha^{1/2}( ds^2+du^2+u^2d{\tilde\psi}^2+
s^2d\chi^2)\cr &+( \alpha^{1/2}- \beta \alpha^{-1/2} ) \left[
{(dx^-)^2\over s^2+u^2}  - {u^4d{\tilde\psi}^2- 2s^2u^2d\chi
d{\tilde\psi} +s^4d{\chi}^2 \over s^2 + u^2 } \right]
}}
where now $\alpha$ becomes \eqn\scaledalph{ \alpha = 1 - ( 1 - \beta)
e^{ -\tilde m (u^2 + s^2) } } Note that $\beta$ remains fixed
and $ 1-\beta $ has the interpretation of momentum  $p_-$ per
unit length. This metric is non-singular. In the limit $\tilde m
\to \infty$ it becomes the metric \PPlimit\ which is singular. Of
course for large $\tilde m$ the metric looks like \PPlimit\ if
$s^2 + u^2 > 1/\tilde m$ where we can approximate $\alpha \sim 1$.

It would be nice to understand why the asymptotic structure of
\PPlimitfull\ (where we can safely set $\alpha =1$) is naively
different
from that of a usual plane wave. For example the transverse space
is no longer  $R^4$ in   \PPlimit .
This deserves further study.

\subsec{Mass gap for the geometry \internapp.}

Let us look at the spectrum of excitation on the background \internapp. For simplicity we will look at the
minimally coupled scalar field, but our results will be true for more general excitations.
Let us remind the reader that for a conical defect metric
with opening angle  $2 \pi \gamma$
the mass gap is $E = 2 \gamma$. Since we argued
above that these conical metrics are not a good approximation to
the metrics we consider we will perform an estimate of  the mass
gap in the metric \internapp.

For simplicity we will look at the modes of the scalar field which are constant in the $\phi,\psi,\chi$
directions. Then looking for the solution in the form $\Phi(r,\theta,t)=e^{-iEt}\Phi(r,\theta)$, we find the
Klein--Gordon equation: \eqn\waveEqn{ {1\over r}\partial_r(r(r^2+\beta)\partial_r \Phi)+{1\over
\sin\theta\cos\theta}
\partial_\theta(\sin\theta\cos\theta\partial_\theta \Phi)+
{E^2\Phi\over (r^2+\beta\cos^2\theta)}
\left[\alpha-{\beta^2\sin^2\theta\over r^2+\beta}\right]=0
}
It is convenient to introduce new coordinates $x=r/\sqrt{\beta}$,
$y=\cos\theta$. Then we find:
\eqn\waveEqnScale{\eqalign{
&{1\over x}\partial_x(x(x^2+1)\partial_x \Phi)+{1\over y}
\partial_y(y(1-y^2)\partial_y \Phi)\cr
&\qquad +{E^2\Phi\over \beta(x^2+y^2)}
\left[\left(1-(1-\beta)\left({1-y^2\over 1+x^2}\right)^m\right)-
\beta{1-y^2\over 1+x^2}\right]=0
}}
Note that for $m=1$ the variables in this equation separate, and in
particular we get spherically symmetric
solutions which are normalizable near $x=0$:
$$
\Phi=(x^2+1)^{-k}F(-{k},1-{k};1;-x^2)
$$
where $E=2k\sqrt{\beta}$. This function in normalizable near infinity if and
only if $k$ is a positive integer. Thus for $m=1$ we have a mass gap
$E=2\sqrt{\beta}$.

Let us now look at the more interesting case when $m > 1$. In
this case the variables in the equation \waveEqnScale~ do not separate, and we
can't find the exact spectrum. But we can get an upper bound on the mass gap
using variational methods. First we rewrite \waveEqnScale~ as a Schroedinger
equation:
$$
(H-E^2V)\Phi=0.
$$
where $V$ is a positive potential ($E^2 V$ comes from the last term in
\waveEqn ).  This eigenvalue
problem is the same as the
 one that arises when we have masses and springs, except that now the
matrices are replaced by operators in a Hilbert space.
Suppose this equation has a spectrum of eigenvalues
$E_k$ with corresponding eigenfunctions $\Phi_k$  obeying
$(H-E^2_k V)\Phi_k =0$.
These will generically
form a complete basis
system in the space of normalizable functions.
Then for any such function we get
$$
\Phi=\sum a_m\Phi_m
$$
We also note that
$$
\langle\Phi_k|(H- E^2V)|\Phi_l\rangle=(E_k^2-E^2)
\langle\Phi_k|V|\Phi_l\rangle=(E_l^2 - E^2)
\langle\Phi_k|V|\Phi_l\rangle
$$
 {}From here we conclude that
$$
\langle\Phi_k|V|\Phi_l\rangle=0,\quad if\quad k\ne l
$$
(as usual, if there is a degeneracy $E_k=E_l$, the above condition gives a choice of a basis).
For a generic function $\Phi$ we get:
$$
\langle\Phi|(H - E^2V)|\Phi\rangle=
\sum ( E_k^2 - E^2 )|a_k|^2\langle\Phi_k|V|\Phi_k\rangle
$$
Since $V$ is positive, $\langle\Phi_m|V|\Phi_m\rangle\ge 0$. To show that
there is an eigenvalue $E_0<E$ it is
sufficient to find a normalizable state $|\Phi\rangle$ such as
$$
\langle\Phi|(H - E^2V)|\Phi\rangle < 0
$$
Let us take a trial function
$$
\Phi={1\over 1+x^2}.
$$
Then taking an average of the left hand side of \waveEqnScale, we find
\eqn\AverRad{\eqalign{
&\langle\Phi|(H - E^2V)|\Phi\rangle
=- {1\over 2}\int_0^\infty dx{\Phi}
\partial_x(x(x^2+1)\partial_x \Phi)\cr
&\qquad - \int_0^\infty xdx{\Phi^2}
{E^2\over\beta}\int_0^1 {ydy\over (x^2+y^2)}
\left[1-(1-\beta)\left({1-y^2\over 1+x^2}\right)^m-
\beta{1-y^2\over x^2+1}\right]\cr
&=\int_0^\infty {udu\over (1+u)^3} - {E^2\over\beta}\left\{
I_0-(1-\beta)I_m-\beta I_1\right\}
}}
Here we introduced the following integral
\eqn\intProbe{\eqalign{
I_k&=\int_0^\infty {xdx\over(1+x^2)^2}
\int_0^1 {ydy\over (x^2+y^2)}
\left({1-y^2\over 1+x^2}\right)^k
=
{1\over 4}\int_0^\infty
{dx\over (x+1)^{k+2}}\int_0^1{(1-y)^kdy\over x+y}\cr
&={1\over 4}\int_0^\infty
{dx\over (x+1)^{k+2}}{1\over x}F(1,1;k+2;-{1\over x})B(1,k+1)\cr
&={1\over 4(k+1)}
}}
This gives
\eqn\ProbeResult{\eqalign{
\langle\Phi|(H- E^2V)|\Phi\rangle={1\over 2} -
{E^2\over 4\beta}\left\{1-
{\beta\over 2}-{1-\beta\over m+1}\right\}
}}
This expression becomes negative  for $E>E_1$, where
\eqn\ProbeGap{
E_1=\sqrt{2\beta}\left\{1-{\beta\over 2}-{1-\beta\over m+1}\right\}^{-1/2}
}
so the mass gap is less than $E_1$. In particular, for all $m\geq 1$
we have
$E_1\le 2\sqrt{\beta}$, so the mass gap is always less than this amount.

\appendix{C}{ No conical defects with arbitrary opening angles}

One can easily write singular solutions with the same  angular momentum
as the solutions we have been considering. The simplest is a
 conical metrics of the form
\eqn\coni{
{ ds^2 \over R^2_{AdS} }  = - (r^2 + \gamma^2) dt^2 +
r^2 d \chi^2 + {d r^2 \over r^2 + \gamma^2}  +
d\theta^2 + \cos^2\theta (d \psi + \gamma d\chi)^2 +
\sin^2 \theta ( d\phi + \gamma d t)^2
}
These metrics have a conical singularity at $ r = (\pi/2 - \theta)=0$.
The singularity has a form which is rather similar to that of
an $A_{N}$ singularity but with an opening angle which is
$2 \pi \gamma$ instead of $2\pi/N$. In addition, if $\gamma^{-1}$
is not an integer there are singularities at $r=0$ and any $\theta$.

When $\gamma^{-1}$ is an integer we can think of the metric \coni\
as arising from a ``supertube'' configuration with $N$ KK monopoles
instead of just one KK monopole. Furthermore, it is possible to
continuously deform the non-singular solutions that we had in this
paper and get to these conical metrics. All we need to do is to
take a profile $F(v)$ which wraps $N$ times around the origin. If
it does not self intersect we will have a smooth metric and as
we take the limit that $F$ is moving on the same circle $N$ times we
get the conical defect metric \coni\ with $\gamma^{-1} =N$.

On the other hand the metrics \coni\ with $\gamma^{-1} \not = N$
should not be allowed from the
 KK monopole point of view since they would
mean that we have fractional KK monopole charge.
In fact this is the reason that the singularity for non-integer
$\gamma^{-1}$
is more extended than for $\gamma^{-1}$ integer. In the former case
there is a fractional ``Dirac string'' coming out of the fractional
KK monopole which is responsible for this singularity.
Despite this strange features one might ask the following question.
Can we find a smooth metric that is arbitrarily close to the metric
\coni\ with non integer $\gamma^{-1}$ ?
When we say that  a metric is { \it very close}  to \coni\ we
mean that the metric is equal to \coni\ up to very small corrections
everywhere except very near the singularity.
Namely, if we pick a $\gamma^{-1}$, say 3/2, then we pick an $\epsilon
$, say $\epsilon = 10^{-6}$, then we want to find a metric which only
differs from \coni\ by terms of order $\epsilon$ once we are at
 $r > \epsilon$.
We will now show that this is {\it not} possible\foot{ So,
for example, it
is futile to try to find the dual description of the conical defect
metrics with arbitrary $\gamma$ \deboeretal , since these metrics
are not a good approximations to anything. It is OK to consider the
ones with integer $\gamma^{-1}$.  }.

Without loss of generality we can assume that the angular momentum is in the direction $J_{12}$ and all other
components vanish. In general the angular momentum of any configuration is characterized by two invariants $J_L^2$
and $J_R^2$ but for conical metrics of the asymptotic form \coni\ we have $J_L^2 = J_R^2$ so that using a rotation
we can always put the angular momentum in the 12 plane. So suppose we have a metric that is very close to the
metric of the conical defect for distances larger than some tiny  distance $\epsilon$. Then the harmonic functions
will be very similar to the harmonic functions that give \coni . The harmonic functions for \coni\ are given by
\ringharm\ except that $\omega$ now obeys $ \omega Q = \gamma R $. Since the harmonic functions are close to each
other the source for the hypothetical non-singular metric should be close to the source of the harmonic functions
in \ringharm . In particular, $f_5$  implies that the source is distributed near a ring in the $12$ plane. So in
the expressions we will find below we will approximate $ F_1^2 + F_2^2 -(F_3^2 + F_4^2) \sim  F_1^2 + F_2^2$, but
we do not make any assumptions about $\dot F_{3,4}^2$.

It is now instructive to consider the large $r$ behavior of the metric. Using \TorMetr\ we can read off all the
harmonic functions of the form \eqss . The leading behavior of such functions is \eqn\leadbeh{\eqalign{
&f_5={Q_5\over x^2}+{2Q_5\langle F_i\rangle x_i\over x^4}+ \langle 4F_iF_j-F^2\delta_{ij}\rangle {x_ix_j\over
x^6}\cr &f_1={Q_5\langle |{\dot G}|^2\rangle\over x^2}+ {2Q_5\langle F_i|{\dot G}|^2\rangle x_i\over x^4}+ \langle
(4F_iF_j-F^2\delta_{ij})|{\dot G}|^2\rangle {x_ix_j\over x^6}, \cr &A_i=-2Q_5\langle{\dot F}_iF_j\rangle {x_j\over
x^4},\quad B_i=-Q_5\eps_{ijkl}\langle{\dot F}_kF_l\rangle {x_j\over x^4},\quad {\cal A}_a=-2Q_5\langle{\dot {\cal
F}}_aF_j\rangle {x_j\over x^4}, }} First let us note that by shifting the origin we can always set $\langle F_i
\rangle =0 $. Then the ten dimensional dilaton will be of the form \eqn\dildencon{ e^{2 \Phi} = { f_1 \over f_5} =
{ Q_1 \over Q_5} ( 1 +2  { x^i \over x^2} { \langle F_i |\dot G|^2 \rangle \over
 {\langle |\dot G |^2 \rangle } } + \cdots )
}
Since this decays very slowly for large $x$ we set its coefficient
to zero.
Similarly, by considering the fields that are excited by the
torus fluctuations we conclude that we also need to set
to zero $\langle {\cal F}_a F_i \rangle =0$.

We have seen above that our metrics will generically have
a particular operator of weight $(1,1)$ with a non-vanishing expectation
value. This will give rise to a deformation of the metric that
can be sensed far away. If we are interested in having a metric
which is very close to the metric of a conical defect then we want
to make the coefficient of this operator as small as possible.
The operator we discussed is an $l=2$ spherical harmonic on
$S^3$ so that its coefficients have the form of a quadrupole
moment ${\cal Q}_{ij}$.
In particular we can look at the following combination:
$$
\sqrt{{\tilde f}_1f_5}-{1\over\sqrt{{\tilde f}_1f_5}}[A_iA_i-B_iB_i]=
{\sqrt{Q_1Q_5}\over x^2}+{\cal Q}_{ij}{x_ix_j\over x^6}+O(x^{-5})
$$
then we notice that the quadrupole moment ${\cal Q}_{ij}$ vanishes for a
conical defect. For a general metric ${\cal Q}_{ij}$ be computed by using
\leadbeh\ and performing a
computation very similar to the one we did near \ntsix,\ \ntseven .
We find
\eqn\coeffoper{
{\cal Q}_{11} + {\cal Q}_{22}
\sim \left \langle [ (F_1^2 + F_2^2) - (F_3^2 + F_4^2 )]
( 1 + {| \dot G|^2 \over \langle  |\dot G|^2 \rangle } )
\right\rangle  - 8 { \langle \dot F_1 F_2 \rangle^2 \over
\langle  |\dot G|^2 \rangle }
}
where expectation values mean averages over $v$ and we used that the
angular momentum is in the 12 plane\foot{Note that
$2\langle F_1 \dot F_2 \rangle = \langle F_1 \dot F_2 -
F_2 \dot F_1 \rangle \sim J_{12} $.}.
We want \coeffoper\  to vanish in order to have a metric close to \coni .
As we argued above we can neglect $ (F_3^2 + F_4^2 )$ relative
to $(F_1^2 + F_2^2)$ in \coeffoper .
It is possible
to show that the result we get after neglecting such a term
is always positive and it only vanishes when the profile is
precisely a ring and the motion has constant velocity.
In order to show that let us multiply all terms in \coeffoper\
by $\langle  |\dot G|^2 \rangle$. Defining
 $F_1 + i F_2 = re^{i\phi} $ we then
find
\eqn\nowcoef{
\left( \langle r^2 \rangle \langle r^2 \dot \phi^2 \rangle - \langle
r^2 \dot \phi \rangle ^2 \right) +
\left(  \langle r^4 \dot \phi^2 \rangle -  \langle
r^2 \dot \phi \rangle ^2 \right) + {\rm other~ terms}
}
 where all other
terms are non negative.
Using the formula $\langle a b\rangle^2
\leq \langle a^2\rangle \langle b^2 \rangle $ (and the equal sign
holds only if $a/b = $constant) we see that all
terms are non negative so that if \nowcoef\ vanishes then all
terms should be zero. Setting the first term in \nowcoef\ to
zero we get that $\dot \phi =$ constant. Setting the second to
zero we get $r=$constant. Setting to zero all other terms in \nowcoef\
we get that $|\dot {\cal F}|^2 = {\dot F}_3^2 + {\dot F}_4^2 =0$.

What we have shown so far is that if the metric is close to the conical
defect then the profile closely tracks a profile with constant
$r$ and $\dot \phi$. Since $\phi$ has to be single valued
this implies that only integer values of $\gamma^{-1}$ are allowed.

We also see that generic chiral primaries with
$J_{L,R}^{NS} < k/2 $ do not produce conical  metrics
\coni\ but the metrics that we have discussed in our paper. Only very
special chiral primaries  produce   metrics close to
\coni\  with integer $\gamma^{-1} $.

\appendix{D}{Evaluation and Expansion of the integrals $I_1^{(n)}(k) , I_2^{(n)}(k)$}

The integrals we defined in \defis
 \eqn\defisapp{ I_1^{(n)}(k)
\equiv {1\over {2\pi}}\int_0^{2\pi}
{{d\alpha}\over{(1+k\cos\alpha)^{n/2}}} ~~~~;~~~~ I_2^{(n)}(k) =
{1\over {2\pi}} \int_0^{2\pi} {{\cos\alpha
d\alpha}\over{(1+k\cos\alpha)^{n/2}}}
}
are not hard to evaluate.
For even $n$ the integrals are
 \eqn\evenI{\eqalign{
& I_1^{(2)}(k)={1\over \sqrt{1-k^2}} ~~,~~ I_1^{(4)}(k)={1\over
{(1-k^2)^{3/2}}} ~~,~~ I_1^{(6)}(k) =
{{(2+k^2)}\over{2(1-k^2)^{5/2}}} \cr & I_2^{(2)}(k)=
-{{(1-\sqrt{1-k^2})} \over {k\sqrt{1-k^2}}} ~, I_2^{(4)}(k)=-{{
k}\over {(1-k^2)^{3/2}}} ~,
 I_2^{(6)}(k)=-{{3 k}\over {2(1-k^2)^{5/2}}}
}}
For odd $n$ , the integrals involve elliptic functions
\eqn\oddI{\eqalign{ 2\pi I_1^{(1)}(k) &=4{1\over
\sqrt{1+k}}K({\sqrt{{2k}\over {1+k}}}) \cr 2\pi I_1^{(3)}(k) &
=4{\sqrt{1+k}\over {1-k^2}}E(\sqrt{{2k}\over{1+k}}) \cr 2\pi
I_1^{(5)}(k)
&={{4\sqrt{1+k}}\over{3(1-k^2)^2}}[-(1-k)K(\sqrt{{2k}\over{1+k}})+4E(\sqrt{{2k}\over{1+k}})]
 \cr 2\pi I_2^{(1)}(k) &= {4\over
{k\sqrt{1+k}}}[(1+k)E(\sqrt{{2k}\over{1+k}})-K(\sqrt{{2k}\over{1+k}})]
\cr  2\pi I_2^{(3)}(k) &= -{{4\sqrt{1+k}}\over
{k(1-k^2)}}[E(\sqrt{{2k}\over{1+k}})-(1-k)K(\sqrt{{2k}\over{1+k}})]
\cr  2\pi I_2^{(5)}(k) &
={{4\sqrt{1+k}}\over{3k(1-k^2)^2}}[-(1+3k^2)E(\sqrt{{2k}\over{1+k}})+(1-k)K(\sqrt{{2k}\over{1+k}})]
\cr }} They also obey the relations: \eqn\relations{
I_1^{(n)}(-k)=I_1^{(n)}(k) ,~~ I_2^{(n)}(k)=-I_2^{(n)}(-k), ~~~
I_2^{(n)}(k)=-{2\over {n-2}}\partial_k I_1^{(n-2)}(k)  . } We
were interested in evaluating  the integrals at $k=-{2as\over
\sigma^2}$ so that in the near ring limit $k\to -1$ where the
functions \evenI\ and \oddI\ are singular. Let us find the
leading contribution near $k =-1$. For $n>1$ we find
\eqn\Dgfas{\eqalign{ I_1^{(n)}(k)&={1\over 2\pi}\int_0^{2\pi}
{d\alpha\over (1+k-2k\sin^2(\alpha/2))^{n/2}}\approx
{\sqrt{2}\over 2\pi}\int_{-\infty}^{\infty}{d\beta\over
(1+k+\beta^2)^{n/2}}\cr &={\sqrt{2}\over
2\pi}(1+k)^{-n/2+1/2}B({1\over 2},{n-1\over 2}),\cr
I_2^{(n)}(k)&\approx {\sqrt{2}\over
2\pi}(1+k)^{-n/2+1/2}B({1\over 2},{n-1\over 2}). }} For $n=1$ one
can extract the leading asymptotic from the elliptic function
\eqn\DeqFive{ I_1^{(1)}=I_2^{(1)}={\sqrt{2}\over 2\pi} \ln
{32\over 1+k} } We also need the expression for \eqn\eqstam{
I_3^{(n)}\equiv (I_1^{(n)})^2-(I_2^{(n)})^2 } and the asymptotics
\Dgfas, \DeqFive\  is  not enough to evaluate it. Nevertheless,
we can rewrite the leading asymptotics of this expression as
\eqn\stama{ I_3^{(n)}\approx 2I_1^{(n)}(I_1^{(n)}-I_2^{(n)}) }
and the problem is reduced to evaluation of the leading behavior
of \eqn\stamb{ I_4^{(n)}\equiv I_1^{(n)}-I_2^{(n)}\approx {1\over
2\pi}\int_0^{2\pi}{2\sin^2(\alpha/2) d\alpha\over
(1+k-2k\sin^2(\alpha/2))^{n/2}} } This integral can be written as
\eqn\stamc{ I_4^{(n)} \approx {2\over n-2}(k+1)^{2-n/2}{\d\over
\d k}\left[(k+1)^{n/2-1}I_1^{(n-2)}\right] } for $n>1$, and the
integrals for $I_4^{(1)}$ and $I_4^{(2)}$ can be evaluated
explicitly. This gives the following asymptotics: \eqn\stamd{
I_4^{(1)}={2\sqrt{2}\over \pi},\quad I_4^{(2)}={1},\quad
I_4^{(3)}={1\over \pi\sqrt{2}}\ln{32\over 1+k},\quad
I_4^{(n)}={B({1\over 2},{n-3\over
2})\over\sqrt{2}\pi}{(1+k)^{3-n}\over n-2}~(n>3) } Using above
expressions we can find the leading asymptotics of $I_3^{(n)}$
\eqn\Idiffev{\eqalign{ & I_3^{(1)}= {8\over \pi^2}(\ln {8a\over
\rho}),\quad I_3^{(2)}={2a\over \rho},\cr & I_3^{(3)}={8a^2\over
\pi^2\rho^2}(\ln{8a\over \rho}),\quad I_3^{(4)}= ({a\over
\rho})^4,\cr & I_3^{(5)}= {64\over 9\pi^2}{a^6\over \rho^6},\quad
I_3^{(6)}= {3\over 4}({a\over \rho})^8 }} where we have used $$
1+k\approx {\rho^2\over 2a^2} $$

\appendix{E}{ The near ring solutions}

In this appendix we expand \DzeroFone\
around the
ring to examine its behavior.

Expanding \DzeroFone~ for small $\rho$ , where $\rho$ is the distance from the ring,  $s=a+\rho\sin\Theta$,
$w=\rho\cos\Theta$, using the expansions of $I_1,I_2$ around $-1$ which appear in Appendix D , we find the
following near-ring metrics for the different dimensions\foot{More rigorously, all of the limits above should be
thought of as scaling limits where $a,Q,\omega$ remain constant and the coordinates scale. For $d=4$ this scaling
is $\rho,~z^i\sim \epsilon^2$ , $y,\phi,t\sim \epsilon$ , $\epsilon\to 0$. then $ds^2\sim \epsilon^3$. For $d\geq
5$ the scaling is $\rho,z^i\sim \epsilon^2$ , $\phi\sim 1$ , $y\sim \epsilon^{-(d-5)}$ , $t\sim
\epsilon^{-2(d-5)}$ and then the metric scales as $ds^2\sim \epsilon^{7-d}$. In these limits, $g_{tt}$ always
scales to zero as $g_{tt}\sim ({\rho\over a})^{3(d-3)/2}\sim \epsilon^{3(d-3)}$. however, as $g_{t\phi}$ remains
finite in the limit, the metrics we obtain are nondegenerate. }
\item{*} $d =3$ \foot{This form of the metric is valid only for $\rho\ll a$ where the $\log$ is strictly positive. For larger
values of $\rho$, one needs to retain more terms in the expansion
of the elliptic functions. A U-dual system of this $d=3$ solution
was recently considered in \Lugo , where it was lifted to an
M-theory solution with zero gauge fields. That solution is
singular, as can be verified by calculating its curvature
invariants. }: \eqn\Dfive{\eqalign{ ds^2 &\approx{1\over
a^2\omega^2}({a\pi\over
Q})^{3/2}(\ln(8a/\rho))^{-1/2}[
{2a\omega Q\over \pi}dtd\phi +{Q\over \pi a}dy^2 +\cr &
+{a^2\omega^2Q\over \pi }({4\over \pi}+{1+a^2\omega^2\over
a\omega^2 Q})d\phi^2] +\cr &+ \sqrt{{Q\over \pi a}\ln {8a\over
\rho}}[d\rho^2+\rho^2d\Theta^2+dz^2_5] }}
\item{*} $d=4$: \eqn\Dsix{\eqalign{ ds^2 &\approx{1\over a^2\omega^2}({2a^2\over
Q})^{3\over2}({\rho\over a})^{1\over 2}[
\omega Qdtd\phi +{Q\over 2a^2}dy^2 +{\omega^2Q^2\over
2}(1+{1+a^2\omega^2\over \omega^2Q})d\phi^2] \cr &+\sqrt{Q\over
2a^2}({a\over \rho})^{1\over 2}[d\rho^2+\rho^2d\Omega_2^2 +dz^2_4]
}}
\item{*} $d=5$:\eqn\Dseven{\eqalign{ ds^2 &\approx{1\over a^2\omega^2}({\pi a^3\over
Q})^{3/2}({\rho\over a})[
{2\omega Q\over
a\pi}d\phi dt + {Q\over a^3\pi}dy^2 +({\omega Q\over \pi a})^2\ln
{8a\over \rho} d\phi^2] +\cr &+ \sqrt{Q\over a^3\pi}({a\over
\rho})[d\rho^2+\rho^2d\Omega_3^2+dz_3^2] }}
\item{*} $d =6$:\eqn\deight{\eqalign{ ds^2 &\approx {1\over a^2\omega^2}({4a^4\over
Q})^{3/2}({\rho\over a})^{3/2}[
{\omega
Q\over 2a^2}dtd\phi + {Q\over 4a^4}dy^2 +({\omega Q\over
4a^2})^2({a\over \rho})d\phi^2] +\cr &+ \sqrt{Q\over 4a^4}({a\over
\rho})^{3/2}[d\rho^2+\rho^2d\Omega_4^2 +dz_2^2] }}
\item{*} $d =7$:\eqn\Dnine{\eqalign{  ds^2 &\approx{1\over a^2\omega^2}({9\pi a^5\over
2Q})^{3/2}({\rho\over a})^2[
{4\omega Q\over 9\pi a^3}dtd\phi +{2Q\over 3\pi a^5 }dy^2
+{2\omega^2Q^2\over  3\pi^2a^6}({a\over \rho})^2d\phi^2] +\cr &+
\sqrt{{2Q\over 3\pi a^5 }}{a^2\over
\rho^2}[d\rho^2+\rho^2d\Omega_5^2+dz_1^2] }}
\item{*} $d =8$:\eqn\dten{\eqalign{ ds^2 &\approx {1\over a^2\omega^2}[{16a^6\over
3Q}]^{3/2}({\rho\over a})^{5/2}[
{3\omega
Q\over 8 a^4} dtd\phi +{3Q\over 16a^6}dy^2 +{3\omega^2Q^2\over
256a^8}({a\over \rho})^3 d\phi^2]+ \cr &+ \sqrt{3Q\over
16a^6}({a\over \rho})^{5/2}[d\rho^2+\rho^2d\Omega_6^2] }}

Looking at these metrics, one can see that only for $d=4$, we
obtain a $g_{\phi\phi}$ which scales with $\rho$ like the other
metric components parallel to the brane. For the other dimensions
$d>4$, we find that $g_{\phi\phi}$ goes to zero much slower than
the other parallel components, as we approach the brane. However,
one must bare in mind that what we should obtain are supergravity
solutions describing a {\it brane  with fluxes on a ring}. The
effects of the curvature evidently affect the metric near the
brane for all $d > 4$. This is related to I.R. phenomena on the
worldvolume theory on the brane.
Whether a solution is singular or not might depend on the U-duality
frame in which it is presented. We did not find any frame where the
solutions we have above for $d\not = 4$ ($d=4$ is U-dual to the D1-D5
system) are non-singular.

\listrefs

\bye